\documentclass[12pt]{article}

\usepackage[usenames]{color}
\usepackage{setspace}

\usepackage{graphicx}
\usepackage{amsmath}
\usepackage{amssymb}
\usepackage{amsthm}

\usepackage{bm}
\usepackage {arydshln}
\usepackage{ulem}

\usepackage{hyperref}
\hypersetup{
setpagesize=false,
 bookmarksnumbered=true,%
 bookmarksopen=true,%
 colorlinks=true,%
 linkcolor=blue,
 citecolor=blue,
}

\newcommand{\MyDecom}[6]{(\repr{#1},\repr{#2})_{#3,#4,#5}^{#6}}
\newcommand{\Eqref}[1]{eq.~(\ref{#1})}
 
\newcommand{\VEV}[1]{\left\langle #1 \right\rangle}

\newcommand{\der}{\partial}

\newcommand{\nn}{\nonumber}

\newcommand{\repr}[1]{{\bf#1}} 
\newcommand{\sub}[1]{_{\rm{#1}}}

\newcommand{\Z}[1]{{\mathbb Z}_#1}
\newcommand{\abs}[1]{\left| #1 \right|}

\newcommand{\cc}[1]{\overline{\bm{#1}}}

\newcommand{\bequ}{\begin{equation}}
\newcommand{\eequ}{\end{equation}}
\newcommand{\beqn}{\begin{eqnarray}}
\newcommand{\eeqn}{\end{eqnarray}}
\newcommand{\bctr}{\begin{center}}
\newcommand{\ectr}{\end{center}}
\newcommand{\bit}{\begin{itemize}}
\newcommand{\eit}{\end{itemize}}
\newcommand{\Ls}{\left(}
\newcommand{\Rs}{\right)}

\newcommand{\half}{{\frac12}}
\allowdisplaybreaks[4]

\usepackage{multirow}
\numberwithin{equation}{section}

\begin{document}

\begin{flushright}
\today
\end{flushright}

\begin{center}

{\LARGE\bf  Models with rank-reducing discrete boundary conditions on $T^2/\Z4$}

\vskip 1.4cm

{\large  
$^{a}$Yoshiharu Kawamura\footnote{e-mail:haru@azusa.shinshu-u.ac.jp},
$^{a}$Eiji Kodaira\footnote{e-mail:kodaira@shinshu-u.ac.jp},
$^{b}$Kentaro Kojima\footnote{e-mail:kojima@artsci.kyushu-u.ac.jp}\\
and\\
$^{c}$Toshifumi Yamashita\footnote{e-mail:tyamashi@aichi-med-u.ac.jp}
}
\\
\vskip 1.0cm
{\it $^a$Department of Physics, Shinshu University, Matsumoto 390-8621, Japan}\\
{\it $^b$ Faculty of Arts and Science, Kyushu University, Fukuoka 819-0395, Japan}\\
{\it $^c$Department of Physics, Aichi Medical University, Nagakute 480-1195, Japan}
\vskip 1.0cm

\begin{abstract}
We study six-dimensional $SU(n)$ gauge models with rank-reducing discrete boundary conditions on the orbifold $T^2/\Z4$,
without and with continuous Wilson line phases. For the latter case, we find that a minimal model can describe the breakdown of the electroweak symmetry based on an $SU(6)$ gauge group.
This model possesses excellent features that two Higgs doublets come from the zero modes of the extra-dimensional gauge field, and the quarks in each generation can be unified into one multiplet, without exotic quarks, as the zero modes of a bulk field in the $\bm{15}$ representation of $SU(6)$.    
There exists a vacuum where the electroweak symmetry is slightly broken by the Hosotani mechanism, with the addition of suitable bulk fields. 
\end{abstract}

\end{center}

\vskip 1.0 cm

%
\section{Introduction}
\label{sec:intro}
%

Gauge theories defined on a higher-dimensional space-time have attractive features in that
the gauge bosons and the Higgs boson can be unified as a higher-dimensional gauge multiplet~\cite{M}, and the Higgs mass splitting between the doublet components and the triplet ones in the grand unified theories (GUTs)~\cite{GUT,SUSYGUT-DG,SUSYGUT-S} can be elegantly realized by orbifolding~\cite{K1,K2,H&N}.  The four-dimensional (4D) chiral fermions can be generated after the compactification on an orbifold.  For instance, the standard model fermions may
arise from zero modes of bulk fields, or may be localized on fixed points of the orbifold.

A remarkable consequence of the idea to unify the gauge and the Higgs boson is that the effective potential of the Higgs field can be finite without supersymmetry (SUSY)~\cite{finiteness-K,finiteness-HI&L}. This is because the Higgs field becomes a pseudo Nambu-Goldstone mode~\cite{finiteness-AC&G}, with respect to spatially separated symmetry breakings.  In other words, it corresponds to the continuous Wilson line phase, and the non-locality ensures the finiteness of the Wilson-line-dependent part of the effective potential, provided that the vacuum remains sufficiently close to the Wilson-line direction. While there are local divergences in subdiagrams~\cite{finiteness-M&Y,finiteness-HMT&Y,divcont}, they are removed only by the lower-loop counter terms without new counter terms~\cite{finiteness-M&Y,finiteness-HMT&Y}. This idea has been applied to the electroweak symmetry breaking (EWSB)~\cite{GHU-KL&Y,GHU-CG&M,GHU-SS&S} in many models including the GUT context~\cite{GHU-HHK&Y,Lim:2007jv,Hosotani:2015hoa,Maru:2019lit}. In addition, it is also used to break the GUT symmetry with~\cite{gGHU-E6} or without~\cite{gGHU,gGHU-Y,gGHU-KT&Y,Kojima:2023mew} rank reduction. In particular, the latter case in $SU(5)$ model shows that the boundary conditions (BCs) for the Higgs mass splitting assumed in refs.~\cite{K1,K2,H&N} can be naturally obtained as the minimum of the effective potential, and its
phenomenological implications have been studied~\cite{gGHU-pheno,gGHU-pheno-NSS&Y}.

The physical symmetry of each model is determined in cooperation 
with BCs of fields and the dynamics of the continuous Wilson line phases, by the Hosotani mechanism~\cite{H1,H2,HHH&K}.  The BCs on orbifolds are represented by the so-called twist matrices, which are elements of transformation groups of the symmetries of the Lagrangian, and classified by equivalence relations due to the gauge symmetries~\cite{HH&K}.  The classification of the twist matrices
was performed on $S^1/\Z 2$~\cite{HH&K} to 
show that each equivalence class has at least one diagonal representative.  Relying on the result, a general form of the effective potential is derived~\cite{generalFormula}.

In our previous work~\cite{Kawamura:2022ecd}, the existence of diagonal representatives in each equivalence class of twist matrices of BCs has been studied in $SU(n)$ or $U(n)$ gauge theories compactified on the orbifolds $T^2/\Z N$ ($N = 2, 3, 4, 6$), under the assumption that the theory has a global $G' = U(n)$ symmetry.
Using constraints, unitary transformations and gauge transformations, it has been examined whether the twist matrices can simultaneously become diagonal or not, and it has been shown that at least one diagonal representative necessarily exists in each equivalence class on $T^2/\Z 2$ and $T^2/\Z 3$, but the twist matrices on $T^2/\Z 4$ and $T^2/\Z 6$ can contain not only diagonal matrices but also non-diagonal $2 \times 2$ ones and non-diagonal $3 \times 3$ and $2 \times 2$ ones, respectively, as a member of block-diagonal submatrices.\footnote{%
Based on our results, the complete classification of equivalent classes has been studied in $SU(n)$ or $U(n)$ gauge theories compactified on $T^2/\Z N$ using trace conservation laws~\cite{T&I1,T&I2}.}
These non-diagonal matrices have discrete parameters, which means that the rank-reducing symmetry breaking can be caused by the discrete Wilson line phases.

In refs.~\cite{S&S,FN&W}, the rank reduction of the gauge group has been studied in orbifold construction.  It is pointed out that the rank is not reduced by the discrete Wilson lines.  In these references, the rank reduction is discussed mainly focusing on the $\Z2$ orbifolds, and the
possibility of $T_1$ (discrete Wilson lines) that do not commute with the rotation $R_0$ has not been studied on other orbifolds. Our previous results
provide counterexamples to this
claim.  Hence, it is interesting to study phenomenological implications on models with the rank-reducing discrete BCs.

A key difference between the six-dimensional (6D) or higher-dimensional models
and the five-dimensional (5D) models is that
6D models allow tadpole terms of the field strength, 
$F_{56}$, on fixed points. 
Then, generally speaking, such tadpole terms receive the loop corrections, which are 
quadratically divergent already at one-loop level~\cite{vonGersdorff:2002us}. 
Since the commutation relation in the field strength is quadratic in the gauge fields, 
the Higgs masses may suffer from the quadratic divergences to ruin the above mentioned 
merit in the scenario where the gauge and the Higgs fields are unified. 
Although the tadpole terms can be forbidden in the $T^2/\Z 2$ by imposing the 
reflection symmetry along one direction of the extra dimension so that 
$F_{56}\to-F_{56}$, there is no discrete symmetry to forbid these terms completely~\cite{GHU-CG&M}, 
and then the bulk matter content is constrained to cancel the divergences in the 
one-loop corrections in ref.~\cite{Scrucca:2003ut}.

In this paper, we study 6D $SU(n)$ gauge models with rank-reducing discrete BCs on $T^2/\Z4$.  In the absence of continuous Wilson line phases,
product group unification models can be constructed based on $SU(7)$ and $SU(8)$ gauge models.
When continuous Wilson line phases are present, we find that a minimal model, namely, an $SU(6)$ gauge model,
can describe the breakdown of the electroweak (EW) symmetry.
In addition, this
model has excellent features that two Higgs doublets appear
from the zero modes of the extra-dimensional gauge field
as in refs.~\cite{Antoniadis:2001cv,Chang:2012iq,Matsumoto:2014ila,Hasegawa:2015vqa,Akamatsu:2023ird},
and the quarks in each generation can be unified into one multiplet, without 
exotic quarks, as the zero modes of a bulk field in
the $\bm{15}$ representation of $SU(6)$ under BCs respecting the EW symmetry.
We show that there exists a vacuum where the
EW symmetry is slightly broken, 
as a minimum of the effective potential for the continuous Wilson line phases, by adding suitable bulk fields, in a toy model.
This EWSB can be naturally understood in terms of the two Higgs doublet model, and mass eigenstates of the neutral Higgs bosons are examined, including
the component corresponding to the nonflat direction, which does not appear in the continuous Wilson line phases.
As for the tadpole   contributions, we simply assume that they are sufficiently small in the present 
analysis. Under this assumption, the vacuum remains close to the Wilson-line direction, and the 
Wilson-line-based analysis is applicable.

The outline of this paper is as follows. In the next section, we briefly review the classification of the BCs on $T^2/\Z4$. 
Using the new possibility of the rank-reducing discrete BCs, we explore a new class of phenomenological models in section~\ref{sec:candidates}. We investigate an $SU(6)$ model with rank-reducing discrete BCs
and continuous Wilson line phases
in section~\ref{Sec:SU(6)GHU}. In section~\ref{sec:EffPot}, 
we study whether the breakdown of the EW symmetry occurs or not via the Hosotani mechanism based on the one-loop effective potential for the continuous Wilson line phases. We describe the breakdown of the EW symmetry using two Higgs doublet model in section~\ref{sec:HeavierHiggs}. 
In the last section, we give conclusions and discussions. Kaluza-Klein (KK) decompositions of fields on $T^2/\Z4$ are shown in appendix \ref{app:KK}. The derivation of the one-loop effective potential for the continuous Wilson line phases is given in appendix \ref{app:epot}.

%
\section{Classification of the boundary conditions on $T^2/\Z4$}
\label{sec:BCsonT2Z4}
%

Based on ref.~\cite{Kawamura:2022ecd}, we
briefly review the classification of the BCs on $T^2/\Z4$.

Let $z$ be a complex coordinate on a two-dimensional (2D) Euclidean space.
We define the translations ${\cal T}_1$ and ${\cal T}_2$ as
\begin{align}\label{ztransdef1}
  {\cal T}_1:z\to z+1, \qquad 
  {\cal T}_2:z\to z+i,
\end{align}
and the $\pi/2$ rotation ${\cal R}_0$ as
\begin{align}\label{idenZn}
  {\cal R}_0:z\to i z.
\end{align}
Then, the orbifold $T^2/{\mathbb Z}_4$ is given by imposing the identification under the translations in eq.~\eqref{ztransdef1} and the rotation in eq.~\eqref{idenZn} on the system:
$z \sim {\cal T}_1z\sim {\cal T}_2z\sim {\cal R}_0z$. 
From eq.~\eqref{idenZn}, we see $({\cal R}_0)^4={\cal I}$, which is the identity operation.

Using ${\cal R}_0$ and ${\cal T}_{1}$, we can define the translations along $i^{m-1}$ $(m=1,\dots,4)$ direction as
\begin{align}
{\cal T}_{m}={\cal R}_0^{m-1}{\cal T}_1{\cal R}_0^{1-m},
  \qquad {\cal T}_{m}:z\to z+i^{m-1}.
\label{2D-Tn}
\end{align}
Since ${\cal T}_{m}$ are translations, they commute with each other and obey $[{\cal T}_{m},{\cal T}_{m'}]=0$ for any pair $(m,m')$.
In addition, as $i$ is the $4$--th root of unity, these translations satisfy the relations:
\begin{align}\label{ztrans-rels}
\prod_{m=1}^4{\cal T}_m={\cal I}, \qquad {\cal T}_1{\cal T}_3
={\cal T}_2{\cal T}_4={\cal I}.
\end{align}
These relations are summarized as $({\cal T}_m {\cal R}_0^p)^{4/p} = {\cal I}$ with integers $p$ and $4/p$~\cite{S&S}. 

On the universal covering space of the orbifold $T^2/{\mathbb Z}_4$, there
are invariant points under ${\cal R}_0$ up to the translations in eq.~\eqref{ztransdef1}, called fixed points.  One can also define the rotations around the fixed points on the orbifold as shown below. If $z_{\rm F}$ is a fixed point, it follows that
\begin{align}\label{fxdptg1}
  z_{\rm F}= i z_{\rm F}+n_1+n_2 i, \qquad n_1,n_2\in{\mathbb Z}. 
\end{align}
Let $z_{{\rm F}}^{(n_1,n_2)}$ be the solutions of the above equation. We obtain the following solutions:
\begin{align}\label{z23fp}
z_{{\rm F}}^{(n_1,n_2)}=\frac{n_1-n_2+(n_1+n_2)i}{2}.
\end{align}
     
Any operation ${\cal T}_1^{n_1}{\cal T}_2^{n_2}{\cal R}_0$ $(n_1,n_2\in{\mathbb Z})$ gives the $\pi/2$ rotation around the fixed point $z_{{\rm F}}^{(n_1,n_2)}$. This is understood because the solution of the equation ${\cal T}_1^{n_1}{\cal T}_2^{n_2}{\cal R}_0z-u=i(z-u)$ for $u$ is nothing but the fixed point defined in eq.~\eqref{fxdptg1}. Thus, the relation $({\cal T}_1^{n_1}{\cal T}_2^{n_2}{\cal R}_0)^4={\cal I}$ is satisfied. Note that there is a ${\mathbb Z}_2$ subgroup in ${\mathbb Z}_4$ and the fixed points under the subgroup ${\mathbb Z}_2$ are given by $z_{{\rm F},2}^{(n_1,n_2)}=(n_1+in_2)/2$. The $\pi$ rotation around $z_{{\rm F},2}^{(n_1,n_2)}$ is given by ${\cal T}_1^{n_1}{\cal T}_2^{n_2}{\cal R}_0^2$, which satisfies $({\cal T}_1^{n_1}{\cal T}_2^{n_2}{\cal R}_0^2)^2={\cal I}$. In addition, four successive rotations around different fixed points can always be expressed by combining the translations as $\prod_{a=1}^4({\cal T}_1^{n_1^{(a)}}{\cal T}_2^{n_2^{(a)}}{\cal R}_0)={\cal T}_1^{n_1'}{\cal T}_2^{n_2'}$, where $n_1^{(a)}$,
$n_2^{(a)}$, $n_1'$ and $n_2'$ are integers.

In the following sections, we consider $SU(n)$ gauge theories on $M^4\times T^2/{\mathbb Z}_4$.  We denote coordinates on $M^4$ and $T^2/{\mathbb Z}_4$ by $x^\mu$ ($\mu=0,1,2,3$) and $z=x^5+ix^6$, respectively.  The translations and the $\pi/2$ rotations accompany non-trivial twists in the representation space of $G'=U(n)$, under which the Lagrangian is supposed to be invariant.  We denote the twist matrices corresponding to these operations by the Italic character symbol, e.g., $R_0$ for ${\cal R}_0$ and $T_m$ for ${\cal T}_m$, which are unitary matrices belonging to the fundamental representation.  These matrices are constrained by the relations satisfied by the corresponding translations and rotations. For example, the relation $[{\cal T}_m,{\cal T}_{m'}]=0$ gives the constraint $[T_m,T_{m'}]=0$ for the twist matrices.

By choosing the twist matrices $R_0$ and $T_1$ as independent ones and using constraints and unitary transformations, they can become block-diagonal forms containing $4 \times 4$, $2 \times 2$ and $1 \times 1$ matrices.  Here, $1 \times 1$ matrices correspond to a diagonal part. The $4\times 4$ submatrices have the form of $r_0$ and $t_1$ presented as
\begin{align}
r_0= \left(
\begin{array}{cccc}
	i &  &  &  \\
	 & -1 &  & \\
	 &  & -i & \\
	 &  &  & 1 
\end{array}
\right),
\qquad
t_1 = e^{i\left(\theta Y + \overline{\theta} Y^{\dagger}\right)} \left(
\begin{array}{cccc}
	1 &  &  &  \\
	 & 1 &  & \\
	 &  & 1 & \\
	 &  &  & 1 
\end{array}
\right),
\label{Z4-t1}
\end{align}
respectively, where $\theta$ is a complex number and $Y$ is a cyclic permutation matrix defined by
\begin{align}
Y \equiv \left(
\begin{array}{cccc}
 & & & 1 \\
1& & &    \\
 &1& &    \\
 & & 1& 
\end{array}
\right).
\label{S}
\end{align}
The $2\times 2$ submatrices have the form of $r'_0$ and $t'_1$ presented as
\begin{align}
r'_0 = i^{n'} \left(
\begin{array}{cc}
	-1 &  0  \\
	0 & 1  
\end{array}
\right),
\qquad
t'_1 = \left(
\begin{array}{cc}
0 & 1 \\
1 & 0 
\end{array}
\right),
\label{Z4-t'1}
\end{align}
where $n'$ is an integer. Combining $r'_0$ with an odd $n'$ and that with an even $n'$, we can make the $4 \times 4$ matrices $r_0$ and $t_1$ such as
\begin{align}
r_0= \left(
\begin{array}{cccc}
	i &  &  &  \\
	 & -1 &  & \\
	 &  & -i & \\
	 &  &  & 1 
\end{array}
\right),
\qquad
t_1 = \left(
\begin{array}{cccc}
0 & 0 & 1 & 0 \\
0 & 0 & 0 & 1 \\
1 & 0 & 0 & 0 \\
0 & 1 & 0 & 0 
\end{array}
\right)
= e^{i\left(i\frac{\pi}{2} Y -i \frac{\pi}{2} Y^{\dagger}\right)} \left(
\begin{array}{cccc}
	1 &  &  &  \\
	 & 1 &  & \\
	 &  & 1 & \\
	 &  &  & 1 
\end{array}
\right),
\label{Z4-t1(a2=1)}
\end{align}
which are diagonalized simultaneously by a suitable gauge transformation, as explained below. Hence, non-paired $r'_0$'s with $n'={\rm odd}$ alone or $n'={\rm even}$ alone remain independent from $r_0$ and $t_1$.

The $4\times4$ submatrices of eq.~\eqref{Z4-t1} can be diagonalized by a suitable gauge transformation, such that
\begin{align}
 \tilde{r}_0 = \varOmega(iz, -i\bar{z})
  r_0 \varOmega^{\dagger}(z, \bar{z}) = r_0,
\qquad
 \tilde{t}_1 = \varOmega(z + 1, \bar{z}+1) t_1 \varOmega^{\dagger}(z, \bar{z})
= (-1)^{\tilde{l}} I,
\label{Gtr.r0t1-prime}
\end{align}
using the gauge transformation function:
\begin{align}
 \varOmega(z, \bar{z}) 
= e^{i\left(\beta z Y + \bar{\beta}\bar{z} Y^{\dagger}\right)},
\label{Gtr.Omega}
\end{align}
with $\displaystyle{\beta= -\theta + \frac{1+i}{2} \pi \tilde{l}}$ including an integer $\tilde{l}$. 
Here, $I$ is the $4\times4$ unit matrix. In a special case with the submatrices of
eq.~\eqref{Z4-t1(a2=1)}, $t_1$ is diagonalized as $\tilde{t}_1 = \varOmega(z + 1, \bar{z}+1) t_1 \varOmega^{\dagger}(z,\bar{z})=I$, keeping $r_0$ the same diagonal one, by using the gauge transformation function $\varOmega(z, \bar{z}) = e^{i\left(-i\frac{\pi}{2} z Y + i\frac{\pi}{2}\bar{z} Y^{\dagger}\right)}$.
The physical symmetries are not determined solely by twist matrices but also by the values of the continuous Wilson line phases, as will be explained below.
In fact, the above gauge transformation also induces a shift of the vacuum expectation values (VEVs) of the gauge fields 
to keep the physical symmetry.
The existence of the diagonal representative of the twist matrices in eq.~\eqref{Z4-t1} means that there is a parameter region of the VEVs of the phases where the rank of the gauge symmetry is not reduced, while the phases generically reduce the rank.

In contrast, the $2\times2$ submatrix $t'_1$ cannot become diagonal ones, keeping $r'_0$ diagonal ones, by the use of a gauge transformation.

In this way, we find that a diagonal representative does not necessarily exist in each equivalence class on $T^2/\Z 4$, because the twist matrices on $T^2/\Z 4$ can contain not only diagonal matrices but also non-diagonal $2 \times 2$ ones as a member of block-diagonal submatrices.

Here, we discuss continuous Wilson line phases, physical symmetries, a reduction of rank and non-diagonal BCs.

The Wilson line phases are non-integrable phases of, for instance, $W_1=e^{ig(\langle A_z \rangle + \langle A_{\bar{z}} \rangle)}T_1$ with the VEVs of the zero modes in $A_z = (A_5 - iA_6)/2$ and $A_{\bar{z}} = (A_5 + iA_6)/2(= A_{z}^\dag)$.  
Here, $A_5$ and $A_6$ are the gauge fields along the coordinates $x^5$ and $x^6$, respectively. The VEV $\langle A_z \rangle$ should change as
\begin{align}
R_0 \langle A_z \rangle R_0^{-1} = i \langle A_z \rangle,
\qquad
R_0 \langle A_{\bar{z}} \rangle R_0^{-1} = -i \langle A_{\bar{z}} \rangle,
\label{Gtr-<Az>}
\end{align}
under the ${\mathbb Z}_4$ rotation.

The physical symmetries are determined in cooperation with BCs of fields and the dynamics of the continuous Wilson line phases, by the Hosotani mechanism~\cite{H1,H2,HHH&K}.  In concrete terms, starting from BCs specified by the twist matrices $(R_0, T_1)$, we calculate the effective potential $V_{\rm eff}$ for the continuous Wilson line phases, which is generated by the quantum corrections.  And then, minimization of $V_{\rm eff}$ tells us the VEV, $\langle A_z \rangle$. After performing the gauge transformation with $\varOmega(z, \bar{z}) = e^{ig(\langle A_z \rangle z + \langle A_{\bar{z}} \rangle \bar{z})}$, $\langle A_z \rangle$ is shifted to $\langle A'_z \rangle = 0$ and the twist matrices are transformed as
\begin{align}
(R_0, T_1) \to (R_0^{\rm sym}, T_1^{\rm sym})
= (\varOmega(iz, -i\bar{z})R_0\varOmega^{\dagger}(z, \bar{z}), 
\varOmega(z+1, \bar{z}+1)T_1\varOmega^{\dagger}(z, \bar{z})).
\label{Gtr-R0-sym}
\end{align}
We note that, based on the definition of continuous Wilson line phases as eigenvalues of $-i\ln W_1$, they are gauge invariant because a change of $\langle A_z \rangle$ is canceled out by that of $T_1$.  The physical gauge symmetry $\mathcal{H}^{\rm sym}$ is spanned by the generators $T^a$ that commute with $(R_0^{\rm sym}, T_1^{\rm sym})$:
\begin{align}
\mathcal{H}^{\rm sym} = \{T^a; [T^a, R_0^{\rm sym}]
=[T^a, T_1^{\rm sym}]=0\}.
\label{Gtr-Ta}
\end{align}

Then, the rank of the unbroken gauge group agrees with the number of Cartan subalgebras that commute with $R_0^{\rm sym}$ and $T_{1}^{\rm sym}$. Thus, the reduction of the rank does not occur if $R_0^{\rm sym}$ and $T_{1}^{\rm sym}$ are diagonal.  In other words, when diagonal $R_0$ and $T_{1}$ belong to the equivalence class, there are symmetry-enhanced points in the parameter space of the VEVs of the continuous Wilson line phases, while the rank is reduced on a generic point. 

In contrast, the reduction of the rank always occurs in the presence of
the $2\times 2$ submatrices in eq.~\eqref{Z4-t'1}
on $T^2/\Z 4$, because they take discrete values and cannot be diagonalized.

In refs.~\cite{S&S,FN&W}, the rank reduction of the gauge group has been studied in orbifold construction.  It is pointed out that the rank is not reduced by the discrete Wilson lines.  In these references, the rank reduction is discussed mainly focusing on the $\Z2$ orbifolds, and the possibility of $T_1$ (discrete Wilson lines) that do not commute with the rotation $R_0$ has not been studied on other orbifolds.  Our previous results provide counterexamples to this claim.  Hence, it is interesting to study phenomenological implications on models with the rank-reducing discrete BCs.\footnote{As another possibility, it is known that BCs related to 
a 't Hooft flux~\cite{tHooft:1979rtg} can reduce the rank~\cite{vonGersdorff:2007uz}.
}

%
\section{Candidates}
\label{sec:candidates}
%

With the newly discovered possibility of the rank-reducing discrete BCs, we explore the new class of phenomenological models based on $SU(n)$ gauge symmetries assuming global $U(n)$ symmetries.

%
\subsection{Without continuous Wilson line phases}
\label{sec:models-w/oWL}
%

First, we examine models without the continuous Wilson line phases.  Since the phases appear
in the $4\times4$ submatrices, we consider combinations of the $1\times1$ and $2\times2$ submatrices in eq.~\eqref{Z4-t'1}.

The minimal model is the $SU(2)$ model, in which the symmetry is completely broken around the compactification scale.  Since there is no
Higgs field, it is not likely appropriate for the EW symmetry.

The next minimal model is an $SU(3)$ model, by adding a $1\times1$ submatrix, with 
\begin{align}
  R_0=\left(
  \begin{array}{cc|c}
    -1&0&\\
    0&1&\\ \hline
    &&1
  \end{array}\right), \qquad
  T_1=\left(
  \begin{array}{cc|c}
    0&1&\\
    1&0&\\ \hline
    &&1
  \end{array}\right),
\label{eq:SU(3)-w/oWL}
\end{align}
by which the $SU(3)$ symmetry is broken down to $U(1)$.  It still does not seem appropriate for the EW symmetry.  We note that the action of $R_0$ itself respects an $SU(2)$ subgroup, and thus an $SU(2)$ symmetry remains on a fixed point, which can be used as a flavor symmetry~\cite{Berezhiani:1985in}.  We note that the matrices in eq.~\eqref{eq:SU(3)-w/oWL} are elements of $U(3)$.
If one changes the signs of the $(3,3)$ entries of the matrices, they become elements of $SU(3)$.

A 4D $SU(2)$ symmetry for the EW symmetry is realized in an $SU(4)$ model, with 
\begin{align}
  R_0=\left(
  \begin{array}{cc|cc}
    -1&0&&\\
    0&1&&\\ \hline
    &&1&\\
    &&&1
  \end{array}\right), \qquad
  T_1=\left(
  \begin{array}{cc|cc}
    0&1&&\\
    1&0&&\\ \hline
    &&1&\\
    &&&1
  \end{array}\right),
\label{eq:SU(4)-w/oWL}
\end{align}
which leads to $SU(2)\times U(1)$ symmetry in the 4D effective theory, although we do not find a significant advantage over the SM. 

In ref.~\cite{Kawamura:2022ecd}, an $SU(7)$ model, by adding the color $SU(3)$ factor, with 
\begin{align}
  R_0=\left(
  \begin{array}{cc|ccccc}
    -1&0&&&&&\\
    0&1&&&&&\\ \hline
    &&1&&&&\\
    &&&1&&&\\
    &&&&1&&\\
    &&&&&1&\\
    &&&&&&1
  \end{array}\right), \qquad
  T_1=\left(
  \begin{array}{cc|ccccc}
    0&1&&&&&\\
    1&0&&&&&\\ \hline
    &&1&&&&\\
    &&&1&&&\\
    &&&&-1&&\\
    &&&&&-1&\\
    &&&&&&-1
  \end{array}\right),
\label{eq:SU(7)-w/oWL}
\end{align}
is shown as an example of the rank-reducing BC without the degrees of the continuous Wilson line phases.  The 4D effective theory has the $SU(3)\times SU(2)\times U(1)^2$ symmetry, which contains extra $U(1)$ factor.

Another possibility to examine in this subsection is to add $2\times2$ matrices. For instance, we may modify the above $SU(7)$ model as 
\begin{align}
  R_0=\left(
  \begin{array}{cc|cc|ccc}
    -1&0&&&&&\\
    0&1&&&&&\\ \hline
    &&-1&0&&&\\
    &&0&1&&&\\ \hline
    &&&&1&&\\
    &&&&&1&\\
    &&&&&&1
  \end{array}\right), \qquad
  T_1=\left(
  \begin{array}{cc|cc|ccc}
    0&1&&&&&\\
    1&0&&&&&\\ \hline
    &&0&1&&&\\
    &&1&0&&&\\ \hline
    &&&&1&&\\
    &&&&&1&\\
    &&&&&&1
  \end{array}\right),
\label{eq:SU(7)-w/oWL-2}
\end{align}
which are rearranged by exchanging the second and third components as
\begin{align}
  R_0=\left(
  \begin{array}{cc|cc|ccc}
    -1&&&&&&\\
    &-1&&&&&\\ \hline
    &&1&&&&\\
    &&&1&&&\\ \hline
    &&&&1&&\\
    &&&&&1&\\
    &&&&&&1
  \end{array}\right), \qquad
  T_1=\left(
  \begin{array}{cc|cc|ccc}
    0&&1&&&&\\
    &0&&1&&&\\ \hline
    1&&0&&&&\\
    &1&&0&&&\\ \hline
    &&&&1&&\\
    &&&&&1&\\
    &&&&&&1
  \end{array}\right).
\label{eq:SU(7)-w/oWL-3}
\end{align}
In this model, the $SU(7)$ symmetry is broken down to $SU(5)\times SU(2)\times U(1)$
symmetry by $R_0$, and further down to the SM symmetry by $T_1$, with no extra $U(1)$ factor.  The breaking pattern by $T_1$ is known as the product group unification~\cite{pgu1,pgu2,pgu3}. We note that this $SU(7)$ model is quite similar to the one studied in ref.~\cite{gGHU-E6} on $S^1/\Z2$, where continuous Wilson line phases cause the latter breaking.  In this sense, this model on $T^2/\Z4$ removes the light degrees of freedom (d.o.f.) of the continuous Wilson line phases from the model on $S^1/\Z2$.

In a similar way, we can construct an $SU(8)$ model for the product group unification with the $SU(5)\times SU(3)\times U(1)$ symmetry, without the light degrees as 
\begin{align}
  R_0=\left(
  \begin{array}{ccc|ccc|cc}
    -1&&&&&&&\\
    &-1&&&&&&\\
    &&-1&&&&&\\ \hline
    &&&1&&&&\\
    &&&&1&&&\\
    &&&&&1&&\\ \hline
    &&&&&&1&\\
    &&&&&&&1
  \end{array}\right), \qquad
  T_1=\left(
  \begin{array}{ccc|ccc|cc}
    0&&&1&&&&\\
    &0&&&1&&&\\
    &&0&&&1&&\\ \hline
    1&&&0&&&&\\
    &1&&&0&&&\\
    &&1&&&0&&\\ \hline
    &&&&&&1&\\
    &&&&&&&1
  \end{array}\right).
\label{eq:SU(8)-w/oWL}
\end{align}
Let us note that $R_0$ and $T_1$ are $U(8)$ elements, which cannot be $SU(8)$ ones by changing the signs of $1\times 1$ submatrices, namely lower-right entries, unlike the $SU(3)$, $SU(4)$ and $SU(7)$ cases.

%
\subsection{With continuous Wilson line phases}
\label{sec:models-w/WL}
%

Next, we examine models with continuous Wilson line phases, in the $4\times4$ submatrices. With the rank-reducing discrete BCs of the $2\times2$ matrix, the minimal model is the $SU(6)$ model with 
\bequ
 R_0=\left(\begin{array}{cc|cccc}
                -1&&&&&\\
                 &1&&&&\\ \hline
                 &&i&&&\\
                 &&&-1&&\\
                 &&&&-i&\\
                 &&&&&1
              \end{array} \right), \qquad
 T_1=\left(\begin{array}{cc|cccc}
                 &1&&&&\\
                 1&&&&&\\ \hline
                 &&1&&&\\
                 &&&1&&\\
                 &&&&1&\\
                 &&&&&1
              \end{array} \right), 
\label{Eq:SU(6)-w/WL}
\eequ
by which the $SU(6)$ symmetry is broken down to $U(1)^4$. The latter BC is a $U(6)$ element.
We note that the rank is reduced from 5 to 4 by the upper-right block of $T_1$ (the $2 \times 2$ matrix $t'_1$).

These BCs admit zero modes for the extra-dimensional components of the gauge fields, $A_z=(A_5-iA_6)/2$ and $A_{\bar z}=A_z^\dagger$, realizing the d.o.f. of the Wilson line phases.  To be more concrete, the zero modes appear for the components that satisfy the relations in eq.~\eqref{Gtr-<Az>} and we can parameterize them, with the normalization of $2\pi R=1$, as
\bequ
g\VEV{A_z}=2\pi\left(\begin{array}{cc|cccc}
                       0&&&&&\\
                        &0&&&&\\ \hline
                        &&&&&\alpha_4\\
                        &&\alpha_1&&&\\
                        &&&\alpha_2&&\\
                        &&&&\alpha_3&
              \end{array} \right)
 =g\VEV{A_{\bar z}}^\dagger. 
\label{Eq:<Az>}
\eequ The remaining $U(1)^4$ symmetry can remove the differences among the phases of $\alpha_j$ $(j=1,\dots,4)$.  In addition, the tree level potential which is proportional to ${\rm Tr}[A_z, A_{\bar z}]^2$ is minimized (vanishing) when the magnitudes $\abs{\alpha_j}$ are also independent of $j$.  We call these directions of the VEVs the flat directions, which correspond to the continuous Wilson line phases. Below, we suppose
that the VEVs are along the flat directions,
unless otherwise noted.  Then, we set $\alpha_j = \alpha = (a-ib)/2$.  Here, we note that the parameter $a$ ($b$) does not correspond to $\VEV{A_5}$ ($\VEV{A_6}$), since the 
matrix \bequ Y=\left(\begin{array}{cc|cccc}
                       0&&&&&\\
                        &0&&&&\\ \hline
                        &&&&&1\\
                        &&1&&&\\
                        &&&1&&\\
                        &&&&1&
              \end{array} \right), 
\label{Eq:Y6}
\eequ
is not hermitian. 

The continuous Wilson line phases are given as, for instance, $W_1=e^{ig(\VEV{A_z}+\VEV{A_{\bar z}})}T_1$, where $g\VEV{A_z} =2\pi\alpha Y$. To deal with the exponential, it is convenient to use the diagonalization of the matrix $Y= UXU^\dagger(=U^\dagger X^\dagger U)$ with 
\bequ
 X=\left(\begin{array}{cc|cccc}
                 0&&&&&\\
                 &0&&&&\\ \hline
                 &&i&&&\\
                 &&&-1&&\\
                 &&&&-i&\\
                 &&&&&1
              \end{array} \right), \qquad
 U=\left(\begin{array}{cc|cccc}
                 -1/\sqrt2&1/\sqrt2&&&&\\
                 1/\sqrt2&1/\sqrt2&&&&\\ \hline
                 &&-i/2&-1/2&i/2&1/2\\
                 &&-1/2&1/2&-1/2&1/2\\
                 &&i/2&-1/2&-i/2&1/2\\
                 &&1/2&1/2&1/2&1/2
              \end{array} \right), 
\label{Eq:X6andU6}
\eequ
where we set the upper left $2\times2$ submatrix of $U$ to diagonalize $T_1$ for later convenience.  Now, $W_1$ is written as $W_1=e^{2\pi i (\alpha Y+\alpha^*Y^\dagger)}T_1=U e^{2\pi i(\alpha X+\alpha^*X^\dagger)}U^\dagger T_1U U^\dagger$, so that
\bequ W_1=U\left(\begin{array}{cc|cccc}
                   -1&&&&&\\
                     &1&&&&\\ \hline
                 &&e^{2\pi ib}&&&\\
                 &&&e^{2\pi i(-a)}&&\\
                 &&&&e^{2\pi i(-b)}&\\
                 &&&&&e^{2\pi ia}
                  \end{array} \right)U^\dagger.
             \label{Eq:W1}
\eequ
We note that $W_1$ is periodic under $a\to a+1$ and $b\to b+1$, and that the remaining $U(1)^4$ symmetry is broken down to $U(1)$, reducing the rank from 4 to 1, for generic $(a, b)$.

In particular, we consider physics around the vacuum with $(a,b)=(0,1/2)$ so that $W_1$ is given as
\bequ
 W_1
      =\left(\begin{array}{cc|c}
                 &1&\\
                 1&&\\ \hline
                 &&\begin{array}{cc:cc}
                        0&&1&\\
                        &0&&1\\ \hdashline
                        1&&0&\\
                        &1&&0
                      \end{array}  
              \end{array} \right). 
\label{Eq:SU(6)GHU-originalW1}
\eequ
We note that the lower-right block of this BC is the same as one in eq.~\eqref{Z4-t1(a2=1)}. It is convenient to permute the index as $\{1,2,3,4,5,6\}\to\{1,3,5,2,6,4\}$ to get
\bequ
 R_0=\left(\begin{array}{cc:cc|cc}
                -1&&&&&\\
                 &-1&&&&\\ \hdashline
                 &&1&&&\\
                 &&&1&&\\ \hline
                 &&&&i&\\
                 &&&&&-i
              \end{array} \right), \qquad
 W_1=\left(\begin{array}{cc:cc|cc}
                 &&1&&&\\
                 &&&1&&\\ \hdashline
                 1&&&&&\\
                 &1&&&&\\ \hline
                 &&&&&1\\
                 &&&&1&
              \end{array} \right). 
\label{Eq:SU(6)GHU-R0W1}
\eequ
We see that the bulk $SU(6)$ symmetry is broken down to $SU(2)\sub{I}\times SU(2)\sub{II}\times U(1)_y\times U(1)_A\times U(1)\sub{III}$ by $R_0$, where the generators of the three $U(1)$'s are respectively defined as
\begin{align}
  T_y&=\frac1{2\sqrt6} t_y, && t_y={\rm diag}(1,1,1,1,-2,-2), \\
  T_A&=\frac1{2\sqrt2} t_A, && t_A={\rm diag}(1,1,-1,-1,0,0), \\
  T_{\rm III}&=\frac1{2} t_{\rm III}, && t_{\rm III}={\rm diag}(0,0,0,0,1,-1).
\label{Eq:SU(6)GHU-U(1)generators}
\end{align}
The generators $T_x$ $(x=y,A,{\rm III})$ are normalized as the fundamental representation of $SU(6)$.  In the following, we refer to eigenvalues of $t_x$ as $U(1)_x$ charges. The continuous Wilson line phases $W_1$ permute the index, and
this permutation is roughly expressed by the exchange of the two $SU(2)$ groups, $SU(2)\sub{I}\leftrightarrow SU(2)\sub{II}$, and the sign flips of the charges of $ U(1)_A$ and $U(1)\sub{III}$. 

It means that on this specific point, the symmetry is enhanced to $SU(2)_D\times U(1)_y$, while it is broken down to $U(1)$ at a generic point away from the specific point. Here, $SU(2)_D$ denotes the diagonal subgroup of $SU(2)\sub{I}\times SU(2)\sub{II}$. This symmetry-breaking pattern may be considered as the EWSB, and the Higgs field should appear as a zero mode of $A_z$, that is, be unified with the 
gauge field. In the following sections, we examine this new type of the $SU(6)$ gauge-Higgs unification model in more detail.

Before closing this section, we show another model in this category, with the $SU(9)$ symmetry, which unifies the above $SU(6)$ symmetry and the color $SU(3)$ symmetry. The BCs are given as
\begin{align}
  R_0=\left(
  \begin{array}{cc|cccc|ccc}
    -1&&&&&&&&\\
    &1&&&&&&&\\ \hline
    &&i&&&&&&\\
    &&&-1&&&&&\\
    &&&&-i&&&&\\
    &&&&&1&&&\\ \hline
    &&&&&&1&&\\
    &&&&&&&1&\\
    &&&&&&&&1
  \end{array}\right), \qquad
  T_1=\left(
  \begin{array}{cc|cccc|ccc}
    &1&&&&&&&\\
    1&&&&&&&&\\ \hline
    &&1&&&&&&\\
    &&&1&&&&&\\
    &&&&1&&&&\\
    &&&&&1&&&\\ \hline
    &&&&&&1&&\\
    &&&&&&&1&\\
    &&&&&&&&1
  \end{array}\right). 
\label{eq:SU(9)-w/WL}
\end{align}
We will examine this extension in a future work.

%
\section{$SU(6)$ gauge-Higgs unification model}
\label{Sec:SU(6)GHU}
%

In the following sections, we investigate the $SU(6)$ model with the BCs in eq.~\eqref{Eq:SU(6)GHU-R0W1}. 

Let us examine the zero modes.  The zero modes are invariant under the $\Z4$ rotation and the translations.  To examine the zero modes, it is convenient to decompose $SU(6)$ multiplets under $SU(2)_{\rm I}\times SU(2)_{\rm II}\times U(1)_y\times U(1)_A\times U(1)_{\rm III}$ and $R_0$.  The action of the $\Z4$ rotation is embedded in $SU(6)$, and we can define charges corresponding to $R_0$ by the generator $t_R=(-2,-2,0,0,1,3)$, where $R_0=e^{2\pi i t_R/4}=i^{t_R}$. We refer to eigenvalues of $t_R$ as $R_0$ charges. Since $R_0$ charges are defined modulo 4, we represent them in the range $[-1,2]$ in the following.  We note that $t_R=-t_y-t_A-t_{\rm III}$ holds.

First, the fundamental representation \repr6 of $SU(6)$ group and its conjugate $\repr{\bar6}$ are given as 
\beqn
 \repr6&=&\Ls\MyDecom{2}{1}{1}{1}{0}{2},\MyDecom{1}{2}{1}{-1}{0}{0},
                    \MyDecom{1}{1}{-2}{0}{1}{1},\MyDecom{1}{1}{-2}{0}{-1}{-1} \Rs^T,
\label{Eq:decom6}\\
 \repr{\bar6}&=&\Ls\MyDecom{2}{1}{-1}{-1}{0}{2},\MyDecom{1}{2}{-1}{1}{0}{0},
                           \MyDecom{1}{1}{2}{0}{-1}{-1},\MyDecom{1}{1}{2}{0}{1}{1} \Rs,
\label{Eq:decom6bar}
\eeqn
where representations/charges of $\Ls SU(2)\sub{I},SU(2)\sub{II}\Rs_{U(1)_y,U(1)_A,U(1)\sub{III}}^{R_0}$ are shown at the indicated positions, and $T$ in \Eqref{Eq:decom6} denotes the transpose.

Generally, in BCs, there is an additional overall factor $\eta_R$ for the $\Z4$ rotation, besides $R_0$, for each multiplet. (We have similar d.o.f. for the translation, $\eta_T$.)  We can choose $\eta_R$ to make any components having the same $R_0$ charge in a multiplet invariant under the $\Z4$ rotation.  For example, $\eta_R=-1=e^{2\pi i\cdot 2/4}$ effectively shifts $R_0$ charges by 2, and then, the component $\MyDecom{2}{1}{1}{1}{0}{2}$ is $\Z4$ invariant. Since this component is not an eigenstate of $W_1$, we see that there is no zero mode.\footnote{%
  It is also understood that, since $R_0W_1=-W_1R_0$, each operation of $W_1$ flips the sign of $R_0$ eigenvalues and shifts $R_0$ charges by 2. Thus, we need an even-rank tensor to obtain the zero modes.}

The higher-dimensional representations can be constructed from the above fundamental ones. An example is the adjoint representation, \repr{35}, accompanied by the trace part \repr1: 
\bequ
 \repr6\times\repr{\bar6}=\repr{35}+\repr1
   = \left(\begin{array}{c:c|cc}
                \MyDecom{3+1}{1}{0}{0}{0}{0}&\MyDecom{2}{2}{0}{2}{0}{2}
                   &\MyDecom{2}{1}{3}{1}{-1}{1}&\MyDecom{2}{1}{3}{1}{1}{-1}\\ \hdashline
                 &\MyDecom{1}{3+1}{0}{0}{0}{0}
                   &\MyDecom{1}{2}{3}{-1}{-1}{-1}&\MyDecom{1}{2}{3}{-1}{1}{1}\\ \hline
                 &&\MyDecom{1}{1}{0}{0}{0}{0}&\MyDecom{1}{1}{0}{0}{2}{2}\\
                 &&&\MyDecom{1}{1}{0}{0}{0}{0}
              \end{array} \right),
\label{Eq:SU(6)GHU-adjoint}
\eequ
where we suppress the lower-left triangle.
For $A_\mu$, $\eta_R=1$ should be taken, and we can show that the linear combination $\MyDecom{3}{1}{0}{0}{0}{0}+ \MyDecom{1}{3}{0}{0}{0}{0}$,
besides the singlets corresponding to $T_y$,
contains the zero mode, which  realizes $SU(2)_D(\times U(1)_y)$ gauge symmetry in the 4D effective theory.
We can understand it from the BCs of the $SU(2)_{\rm I}$ gauge boson $A_{\mu}^{({\rm I})}$
and the $SU(2)_{\rm II}$ gauge boson $A_{\mu}^{({\rm II})}$ given by
\beqn
&&A_{\mu}^{({\rm I})}(x, iz, -i\bar{z}) = A_{\mu}^{({\rm I})}(x, z, \bar{z}),\qquad
A_{\mu}^{({\rm II})}(x, iz, -i\bar{z}) = A_{\mu}^{({\rm II})}(x, z, \bar{z}), 
\label{Eq:SU(6)Amu-R0}\\
&&A_{\mu}^{({\rm I})}(x, z+1, \bar{z}+1) = A_{\mu}^{({\rm II})}(x, z, \bar{z}),\qquad
A_{\mu}^{({\rm II})}(x, z+1, \bar{z}+1) = A_{\mu}^{({\rm I})}(x, z, \bar{z}), 
\label{Eq:SU(6)Amu-W1}
\eeqn
under the $\Z 4$ rotation and the translation represented by the twist matrices
in \Eqref{Eq:SU(6)GHU-R0W1}. 
They are rewritten as
\beqn
A_{\mu}^{(\pm)}(x, iz, -i\bar{z}) = A_{\mu}^{(\pm)}(x, z, \bar{z}),\qquad
A_{\mu}^{(\pm)}(x, z+1, \bar{z}+1) = \pm A_{\mu}^{(\pm)}(x, z, \bar{z}),
\label{Eq:SU(6)Amu(pm)-BC}
\eeqn
where $A_{\mu}^{(\pm)} \equiv (A_{\mu}^{({\rm I})} \pm A_{\mu}^{({\rm II})})/\sqrt{2}$.
For $A_z$ ($A_{\bar z}$), $\eta_R=-i$ ($i$) and we see that the following two linear combinations
\beqn
  &&\MyDecom{2}{1}{3}{1}{-1}{1} + \MyDecom{1}{2}{3}{-1}{1}{1}, 
\label{Eq:SU(6)GHU-Hu}\\
  &&\Ls \MyDecom{2}{1}{3}{1}{1}{-1} + \MyDecom{1}{2}{3}{-1}{-1}{-1}\Rs^*,
\label{Eq:SU(6)GHU-Hd}
\eeqn 
has the zero modes, where the latter comes from the suppressed lower-left triangle in eq.~\eqref{Eq:SU(6)GHU-adjoint}. 

This means that the 4D effective theory of this model is a two Higgs doublet model. We note that, as far as the VEVs of the doublets are small enough compared to the compactification scale, their directions need not be the flat ones, where the tree level potential vanishes. At the tree level, the quadratic terms are vanishing and the quartic terms come from the commutation relations, which are identical to the SUSY D-term contributions. The quadratic terms are generated via loop effects and thus their naive scale is suppressed by the loop factor from the compactification scale. 

From the $U(1)_y$ charges of the Higgs doublets, we find the normalization of the hypercharge $Q_Y$ as $T_y=\frac1{2\sqrt6} 6Q_Y$, whose coefficient $\sqrt{3/2}$ is larger than the corresponding one $\sqrt{3/5}$ in the usual $SU(5)$ GUTs, which means that the unified gauge coupling expected from the observed hypercharge coupling is suppressed.  Let $g_6$ be the 4D gauge coupling constant of the bulk $SU(6)$.
For the $U(1)$ hypercharge and the $SU(2)$ gauge couplings in the SM gauge symmetry, we denote them by $g_Y$ and $g_w$, respectively. We find $g_Y=\sqrt{3/2}g_6$ and $g_w=g_6/\sqrt{2}$, and a large value of the Weinberg angle $\sin\theta_W=\sqrt{3}/2\simeq 0.87$ around the compactification scale. To obtain a realistic low-energy effective theory, effects from boundary operators, renormalization groups, and/or mixing with an additional $U(1)$ gauge symmetry must be included.
For instance, the gauge coupling $g'$ for the additional $U(1)$ can be used to tune the Weinberg angle~\cite{GHU-SS&S,Antoniadis:2001cv}, while the additional degree of the gauge boson can be made massive to decouple from the low energy effective theory.

%
\subsection{Zero modes of matter fields}
\label{Sec:zeromodes-matter}
%

The next smallest representations are the anti-symmetric/symmetric rank-2 tensors: $\repr6\times\repr6 = \repr{15}_A + \repr{21}_S$, with
\bequ
  \repr{15}/\repr{21}
   = \left(\begin{array}{c:c|cc}
                \MyDecom{1/3}{1}{2}{2}{0}{0}&\MyDecom{2}{2}{2}{0}{0}{2}
                   &\MyDecom{2}{1}{-1}{1}{1}{-1}&\MyDecom{2}{1}{-1}{1}{-1}{1}\\ \hdashline
                 &\MyDecom{1}{1/3}{2}{-2}{0}{0}
                   &\MyDecom{1}{2}{-1}{-1}{1}{1}&\MyDecom{1}{2}{-1}{-1}{-1}{-1}\\ \hline
                 &&0/\MyDecom{1}{1}{-4}{0}{2}{2}&\MyDecom{1}{1}{-4}{0}{0}{0}\\
                 &&&0/\MyDecom{1}{1}{-4}{0}{-2}{2}
              \end{array} \right),
\label{Eq:SU(6)GHU-rank2}
\eequ
where the suppressed lower-left triangle is given from the upper-right triangle so that it becomes anti-symmetric/symmetric. 

We see that a \repr{15}
left-handed (right-handed) 
fermion with $\eta_R=1$ ($i$) and $\eta_T=-1$ can be used for $D^c$ and $U^c$ ($Q$).  Since $\eta_R$ of the right-handed component is different from that of the left-handed one by factor $i$ (or $-i$) in a 6D chiral fermion (depending on the 6D chirality), we see that the quarks in each generation can be unified into one multiplet, 
without exotics.  It is in contrast to many gauge-Higgs unification models where two multiplets are necessary.  It is also notable that the quark hypercharges can be realized without an additional $U(1)$ symmetry.

As for the leptons, we need a higher-rank tensor, more than two, to realize the right-handed electron with hypercharge $1$.  From Table~\ref{Table:DecomRank3}, we can see that \repr{56} and \repr{70} may contain the 3 lepton multiplets which are expected to have Yukawa couplings, while they are not eigenstates of $W_1$.  To realize the leptons as zero modes, we would need to examine rank-4 tensors. Another option is to introduce the SM fermions as brane fields and to use the bulk fields as messenger fields for the EWSB~\cite{GHU-CG&M,GHU-SS&S,Maru:2019lit}.

In this bulk sector with the BCs, the discrete symmetry with respect to the exchange between two Higgs doublets remains.
The brane fermions would be helpful to distinguish $H_u$ and $H_d$, or to be more concrete, to realize the difference between the up-type and the down-type quark masses.  Since the gauge symmetry on a boundary is broken, we can put fields in an incomplete multiplet of the $SU(6)$. For example, at the origin, the symmetry is broken down to $SU(2)\sub{I}\times SU(2)\sub{II}\times U(1)_y\times U(1)_A\times U(1)\sub{III}$, we may put fermions in $\MyDecom{1}{1}{-4}{0}{0}{0}$, $\MyDecom{1}{1}{2}{2}{0}{0}$ and $\MyDecom{2}{1}{1}{-1}{-1}{-1}$ for $U^c$, $D^c$ and $Q$ respectively, as independent fields.  These would have bulk-brane mixing mass terms with a bulk \repr{15} fermions to couple with the EWSB, with independent mass parameters. 

\begin{table}[t]
 \begin{center}
  \caption{decompositions of rank-3 tensors}
  \label{Table:DecomRank3}
  \begin{tabular}{ccc}\hline
    \repr{56} & \repr{70} & \repr{20} \\ \hline
     $\MyDecom{4}{1}{3}{3}{0}{2} + \MyDecom{1}{4}{3}{-3}{0}{0}$
    &$\MyDecom{2}{1}{3}{3}{0}{2} + \MyDecom{1}{2}{3}{-3}{0}{0}$
    &$-$\\
     $\MyDecom{3}{2}{3}{1}{0}{0}+\MyDecom{2}{3}{3}{-1}{0}{2}$
    &$\MyDecom{3+1}{2}{3}{1}{0}{0}+\MyDecom{2}{3+1}{3}{-1}{0}{2}$
    &$\MyDecom{1}{2}{3}{1}{0}{0}+\MyDecom{2}{1}{3}{-1}{0}{2}$\\
     $\MyDecom{3}{1}{0}{2}{\pm1}{\pm 1}+\MyDecom{1}{3}{0}{-2}{\mp1}{\mp 1}$
    &$\MyDecom{3+1}{1}{0}{2}{\pm1}{\pm 1}+\MyDecom{1}{3+1}{0}{-2}{\mp1}{\mp 1}$
    &$\MyDecom{1}{1}{0}{2}{\pm1}{\pm 1}+\MyDecom{1}{1}{0}{-2}{\mp1}{\mp 1}$\\
     $\MyDecom{2}{2}{0}{0}{1}{1}+\MyDecom{2}{2}{0}{0}{-1}{-1}$
    &$2\times\Ls\MyDecom{2}{2}{0}{0}{1}{1}+\MyDecom{2}{2}{0}{0}{-1}{-1}\Rs$
    &$\MyDecom{2}{2}{0}{0}{1}{1}+\MyDecom{2}{2}{0}{0}{-1}{-1}$\\
     $\MyDecom{2}{1}{-3}{1}{\pm2}{0}+\MyDecom{1}{2}{-3}{-1}{\mp1}{2}$
    &$\MyDecom{2}{1}{-3}{1}{\pm2}{0}+\MyDecom{1}{2}{-3}{-1}{\mp1}{2}$
    &$-$\\
     $\MyDecom{2}{1}{-3}{1}{0}{2}+\MyDecom{1}{2}{-3}{-1}{0}{0}$
    &$2\times\Ls\MyDecom{2}{1}{-3}{1}{0}{2}+\MyDecom{1}{2}{-3}{-1}{0}{0}\Rs$
    &$\MyDecom{2}{1}{-3}{1}{0}{2}+\MyDecom{1}{2}{-3}{-1}{0}{0}$\\
     $\MyDecom{1}{1}{-6}{0}{3}{-1}+\MyDecom{1}{1}{-6}{0}{-3}{1}$
    &$-$
    &$-$\\
     $\MyDecom{1}{1}{-6}{0}{1}{1}+\MyDecom{1}{1}{-6}{0}{-1}{-1}$
    &$\MyDecom{1}{1}{-6}{0}{1}{1}+\MyDecom{1}{1}{-6}{0}{-1}{-1}$
    &$-$\\
    \hline
  \end{tabular}
 \end{center}
\end{table}

%
\section{Effective potential}
\label{sec:EffPot}
%

The next question is whether the vacuum with $(a,b)=(0,1/2)$ is selected, and
whether the EWSB actually occurs.  To answer this question, we should calculate the effective potential for the continuous Wilson line phases, which are parameterized by $a$ and $b$. 

It is convenient to move to the basis where $W_1$ is diagonal for any $a$ and $b$
to find the coupling with the continuous Wilson line phases. This basis is obtained from the original basis with $W_1$ given in \Eqref{Eq:W1} by the unitary transformation with $U^\dag$, and we get
\bequ R_0=\left(\begin{array}{cc|cccc}
                  &1&&&&\\
                  1&&&&&\\ \hline
                  &&&1&&\\
                  &&&&1&\\
                  &&&&&1\\
                  &&1&&&\\
          \end{array} \right),
      \qquad
      W_1=e^{2\pi i Q}, \qquad
      Q=\left(\begin{array}{cc|cccc}
                1/2&&&&&\\
                   &0&&&&\\ \hline
                   &&b&&&\\
                   &&&-a&&\\
                   &&&&-b&\\
                   &&&&&a
              \end{array} \right).  
\label{Eq:DiagonalW1}
\eequ
We note that the continuous Wilson line phases exist in the generators of the lower-right $SU(4)$ subgroup, which we call $SU(4)_{WL}$, and thus it couples with the $SU(4)_{WL}$ nonsinglets. 

The eigenstates of $W_1$, i.e. the translation eigenstates, are permuted cyclically by $R_0$,
namely, the $\mathbb Z_4$ rotation.  As shown in appendix~\ref{app:KK}, the permutation forms quartets, doublets or singlets, among which only the quartets couple with the continuous Wilson line phases.

We now examine fields $\phi_k$ ($k=1,\dots,4$) that form a quartet as 
\bequ
 \phi_k(iz)=e^{2\pi{ i \gamma_R}}\phi_{k+1}(z), \qquad
 \phi_k(z+1)=e^{2\pi{ i \beta_T}}e^{2\pi iq_k}\phi_k(z), 
\label{Eq:quartet}
\eequ
where we use the convention $\phi_{k+4}=\phi_k$.  Here, $\eta_R$ and $\eta_T$, which are additional overall factors mentioned in section~\ref{Sec:SU(6)GHU}, are parametrized by $\gamma_R$ and $\beta_T$.  As discussed in appendix~\ref{app:KK}, $4\gamma_R\in \mathbb Z$, $q_k+q_{k+2}\in \mathbb Z$ and $2\beta_T\in \mathbb Z$ hold, and their KK masses are summarized as
\bequ
(M^2\sub{KK})^{(n_1,n_2)}=\frac1{R^2}[(n_1+q_1+\beta_T)^2+(n_2+q_2+\beta_T)^2],
\qquad n_1,n_2\in\mathbb Z,
\label{Eq:KKmass}
\eequ
where we write the compactification radius $R$ explicitly, instead of taking $2\pi R=1$.  As usual in the Hosotani mechanism on an orbifold, although the orbifold projection reduces the d.o.f. of each field, the KK masses of the fields in a multiplet that couples to the continuous Wilson line phases recover the whole momentum space in total. 
We see that the periodicity of the continuous Wilson line phases, which is in this case represented by the integer shift $q_k\to q_k+1$, is then ensured.

Once the KK masses are obtained, we can calculate the effective potential for the continuous Wilson line phases. Although each contribution is divergent, the Poisson resummation formula reveals that the divergences are gathered only in the constant term, which is independent of the continuous Wilson line phases, at the one-loop
level.\footnote{%
  At the higher-loop level, divergences in subdiagrams may appear~\cite{finiteness-M&Y,finiteness-HMT&Y,divcont}.  These divergences are removed by the lower-loop counter terms and new counter terms are not necessary~\cite{finiteness-M&Y,finiteness-HMT&Y}.}
Removing the divergent constant term
from eq.~\eqref{effpotform1}, we find the contribution from a real bosonic d.o.f. of a quartet to the effective potential is written as
\bequ
{\cal V}^{[\beta_T]}(q_1,q_2)=
-\frac1{16\pi^7R^4}\sum_{(w_1,w_2)\neq(0,0)} \frac{\cos(2\pi(w_1
  (q_1+\beta_T)+w_2(q_2+\beta_T)))}{(w_1^2+w_2^2)^3}, 
\label{Eq:V-quartet}
\eequ
where $w_1$ and $w_2$ are integers.
The details of the derivation are provided in appendix~\ref{app:epot}. We note that this function is symmetric under $q_1\to q_1+1$, $q_2\to q_2+1$, $q_1\to-q_1$, $q_2\to-q_2$ and $q_1\leftrightarrow q_2$, respectively.  
We also mention that it is independent of the parameter $\gamma_R$, namely $\eta_R$, while it depends on $\eta_T$ as ${\cal V}^{[\beta_T]}(q_1,q_2)={\cal V}^{[0]}(q_1+\beta_T,q_2+\beta_T)$.

%
\subsection{Each representation}
\label{sec:EffPot-repr}
%

Then, we derive the contribution from a real bosonic d.o.f. of some $SU(6)$ representations.  Since the quartet contribution is derived, it is enough to find the quartets with the charges $q_1$ and $q_2$, which are eigenvalues of the charge matrix $Q$ in \Eqref{Eq:DiagonalW1} in the representation. 

We note that the twist matrix $R_0$ also permutes the first two components, and thus that it is helpful to decompose the representation in terms of $SU(4)_{WL}\times SU(2)$ subgroup. In addition, the upper-left $2\times 2$ submatrix of $Q$ is not traceless, and thus we enlarge the $SU(2)$ symmetry to $U(2)=SU(2)\times U(1)$ symmetry. The decompositions are given as
\beqn
\repr6\, &\to&(\repr4,\repr1_0)+\cdots,\\
\repr{15}&\to&(\repr6,\repr1_0)+(\repr4,\repr2_1)+\cdots,\\
\repr{21}&\to&(\repr{10},\repr1_0)+(\repr4,\repr2_1)+\cdots,\\
\repr{35}&\to&(\repr{15},\repr1_0)+(\repr4,\repr2_{-1})+(\repr{\cc4},\repr2_1)+\cdots,\\
\repr{56}&\to&(\repr{\cc{20''}},\repr1_0)+(\repr{10},\repr2_1)+(\repr{4},\repr3_2)+\cdots,\\
\repr{70}&\to&(\repr{\cc{20}},\repr1_0)+(\repr{10},\repr2_1)+(\repr{6},\repr2_1)+(\repr{4},\repr3_2)+(\repr{4},\repr1_2)+\cdots,\\
\repr{20}&\to&(\repr{\cc4},\repr1_0)+(\repr6,\repr2_1)+(\repr4,\repr1_2),
\eeqn
where we suppress the $SU(4)_{WL}$ singlets, which do not couple to the continuous Wilson line phases. Here, the Dynkin labels of the two 20-dimensional representations of the $SU(4)$ group are $(011)$ for $\repr{20}$ and $(003)$ for $\repr{20''}$, in accordance with refs.~\cite{slansky,yamatsu}.

The $SU(4)_{WL}$ fundamental representation, $\repr4$, acts as a quartet, which contributes to the charges $(q_1,q_2)$ by $(b,-a)$.  Note that, thanks to the symmetry of the function ${\cal V}^{[\beta_T]}$ in \Eqref{Eq:V-quartet}, there are equivalence relations for the contribution, for example, $(b,-a)\sim(b,a)\sim(a,b)$. 

Its conjugate representation, $\cc4$, also acts as a quartet with the charge contribution $-(b,-a)\sim(a,b)$, which is equivalent with the one of $\repr4$. This holds for any representation. 

The antisymmetric rank-2 tensor, $\repr6$, contains one quartet with the charge contribution $(b-a,-a-b)\sim(a+b,a-b)$.  It is understood by an explicit representation, $\phi_{ij}$ ($i,j=1,\dots,4$): $(\phi_{12},\phi_{23},\phi_{34},\phi_{41})$ acts as a quartet, while the remaining $(\phi_{13},\phi_{24})$ does as a doublet, which does not couple with the continuous Wilson line phases.

The symmetric rank-2 tensor, $\repr{10}$, contains two quartets with the contributions $(2a,2b)$ and $(a+b,a-b)$, which are $(\phi_{11},\phi_{22},\phi_{33},\phi_{44})$ and $(\phi_{12},\phi_{23},\phi_{34},\phi_{41})$ in the explicit representation, respectively.  We denote it as $\repr{10}\ni(2a,2b)+(a+b,a-b)$ for short.

In a similar way, we can find quartets with the contributions in the adjoint representation, $\repr{15}$, the completely symmetric rank-3 tensor, $\repr{20''}$, and the partly antisymmetric rank-3 tensor, $\repr{20}$. They are summarized as 
\beqn
\repr4\, &\ni&(a,b),\\
\repr6\, &\ni&(a+b,a-b),\\
\repr{10}&\ni&(2a,2b)+(a+b,a-b),\\
\repr{15}&\ni&(2a,2b)+2\times(a+b,a-b),\\
\repr{20''}&\ni&(3a,3b)+(2a+b,a-2b)+(2a-b,a+2b)+2\times(a,b),\\
\repr{20}&\ni&(2a+b,a-2b)+(2a-b,a+2b)+3\times(a,b), 
\eeqn
where for instance $3\times(a,b)$ denotes there are three quartets with the contribution $(a,b)$, instead of one quartet with $3(a,b)=(3a,3b)$. We remind the readers that these are the representations of $SU(4)$ and should not be confused with those of $SU(6)$. 

Now, the effects of the $SU(4)_{WL}$ are understood.  Next, we examine those of the $U(2)$.  Let us first consider $(\repr4,\repr2_1)$ in the decomposition of \repr{15}.  It is easy to understand that there are two \repr4, or quartets.  Their contributions are shifted by the effect of $\repr2_1$.  It is again understood by an explicit representation, $\phi_{i,\alpha}$ ($i=1,2,3,4$ and $\alpha=1,2$): $(\phi_{1,1},\phi_{2,2},\phi_{3,1},\phi_{4,2})$ and $(\phi_{1,2},\phi_{2,1},\phi_{3,2},\phi_{4,1})$ act as quartets, with the charge contributions $(b+1/2,-a)\sim(a,b+1/2)$ and $(b,-a+1/2)\sim(a+1/2,b)$, respectively.  Namely, there are two copies of the quartets, but with different charge shifts.  We denote it as $\repr2_1\ni(0,1/2)+(1/2,0)$ for short.

The other representations of $U(2)$ are understood in a similar way, and summarized as 
\beqn
\repr1_0&\ni&(0,0),\\
\repr2_1&\ni&(0,1/2)+(1/2,0),\\
\repr3_2&\ni&2\times(0,0)+(1/2,1/2),\\
\repr1_2&\ni&(1/2,1/2), 
\eeqn
where $2\times(0,0)$ denotes there are two copies of the quartets with no shifts, as before. 

In the end, we find the contributions from a real bosonic d.o.f. of the representations as 
\beqn
{\cal V}^{[\beta_T]}_\repr6(a,b)&=&{\cal V}^{[\beta_T]}(a,b), 
\label{Eq:V-6} \\
{\cal V}^{[\beta_T]}_\repr{15}(a,b)&=&{\cal V}^{[\beta_T]}(a+b,a-b)+{\cal V}^{[\beta_T]}(a,b+1/2)+{\cal V}^{[\beta_T]}(a+1/2,b), 
\label{Eq:V-15} \\
{\cal V}^{[\beta_T]}_\repr{21}(a,b)&=&{\cal V}^{[\beta_T]}(2a,2b)+{\cal V}^{[\beta_T]}(a+b,a-b)+{\cal V}^{[\beta_T]}(a,b+1/2)+{\cal V}^{[\beta_T]}(a+1/2,b), 
\label{Eq:V-21} \\
{\cal V}^{[\beta_T]}_\repr{35}(a,b)&=&{\cal V}^{[\beta_T]}(2a,2b)+2{\cal V}^{[\beta_T]}(a+b,a-b)+2{\cal V}^{[\beta_T]}(a,b+1/2)+2{\cal V}^{[\beta_T]}(a+1/2,b), 
\notag \\ \label{Eq:V-35} &&\\
{\cal V}^{[\beta_T]}_\repr{56}(a,b)&=&{\cal V}^{[\beta_T]}(3a,3b)+{\cal V}^{[\beta_T]}(2a+b,a-2b)+{\cal V}^{[\beta_T]}(2a-b,a+2b) \nn\\
                       &&+{\cal V}^{[\beta_T]}(2a,2b+1/2)+{\cal V}^{[\beta_T]}(2a+1/2,2b) \nn\\
                       &&+{\cal V}^{[\beta_T]}(a+b,a-b+1/2)+{\cal V}^{[\beta_T]}(a+b+1/2,a-b) \nn\\
                       &&+4{\cal V}^{[\beta_T]}(a,b)+{\cal V}^{[\beta_T]}(a+1/2,b+1/2), 
\label{Eq:V-56} \\
{\cal V}^{[\beta_T]}_\repr{70}(a,b)&=&{\cal V}^{[\beta_T]}(2a+b,a-2b)+{\cal V}^{[\beta_T]}(2a-b,a+2b) \nn\\
                       &&+{\cal V}^{[\beta_T]}(2a,2b+1/2)+{\cal V}^{[\beta_T]}(2a+1/2,2b) \nn\\
                       &&+2{\cal V}^{[\beta_T]}(a+b,a-b+1/2)+2{\cal V}^{[\beta_T]}(a+b+1/2,a-b) \nn\\
                       &&+5{\cal V}^{[\beta_T]}(a,b)+2{\cal V}^{[\beta_T]}(a+1/2,b+1/2), 
\label{Eq:V-70} \\
{\cal V}^{[\beta_T]}_\repr{20}(a,b)&=&{\cal V}^{[\beta_T]}(a+b,a-b+1/2)+{\cal V}^{[\beta_T]}(a+b+1/2,a-b) \nn\\
                       &&+{\cal V}^{[\beta_T]}(a,b)+{\cal V}^{[\beta_T]}(a+1/2,b+1/2).
\label{Eq:V-20} 
\eeqn
We note that ${\cal V}^{[1/2]}_{\cal{R}}(a,b)={\cal V}^{[0]}_{\cal{R}}(a+1/2,b+1/2)$ does not hold except for ${\cal{R}}=\repr6$. In particular, ${\cal V}^{[0]}_\repr{20}(a,b)={\cal V}^{[1/2]}_\repr{20}(a,b)$ holds.
The total effective potential is given by the summation of the above contributions over the bulk fields, with weight factors of the numbers of d.o.f. and sign factors for fermions. For example, the weight factors are $2$, $4$ and $-4$ for complex scalars, the gauge field and 6D chiral fermions, respectively. 

We note that the effective potential is calculated only for the flat directions.  In addition, the effects of the mass terms, for example the bulk masses nor the bulk-boundary mixing masses, are not included.  As discussed later in section \ref{sec:HeavierHiggs}, the effects are generically important and our calculation should be understood just as
a proof of the existence of a vacuum with small breaking of the EW symmetry.

%
\subsection{Numerical calculation}
\label{sec:NumCalc}
%

Here, we show some results of the numerical evaluations. In our calculation, we cutoff the summation over $w_1$ and $w_2$ in \Eqref{Eq:V-quartet} at $\abs{w_i}\leq w_{\rm cut}=100$ ($i=1,2$).

\begin{figure}[]
\centering
  \includegraphics[width=3.9cm,clip]{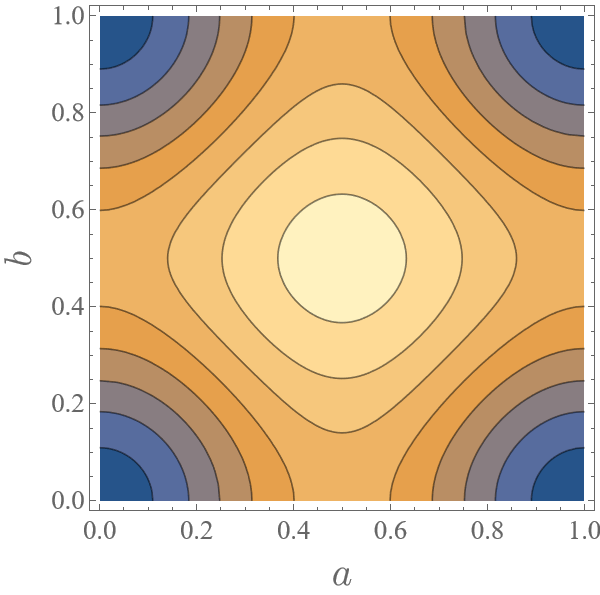} \,
  \includegraphics[width=3.9cm,clip]{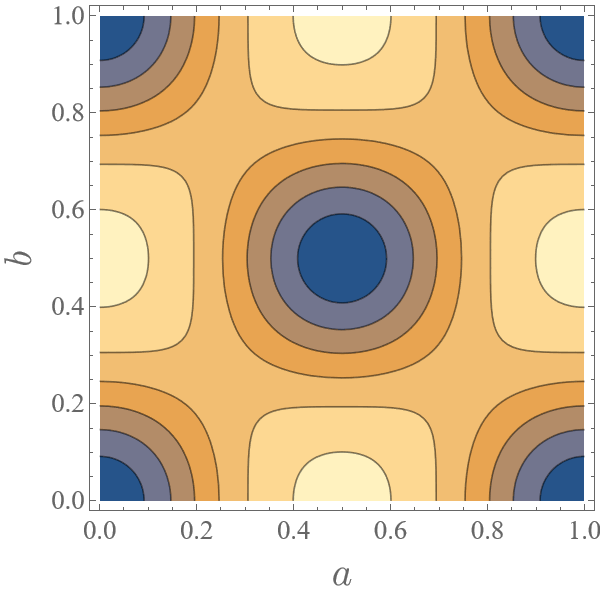} \,
  \includegraphics[width=3.9cm,clip]{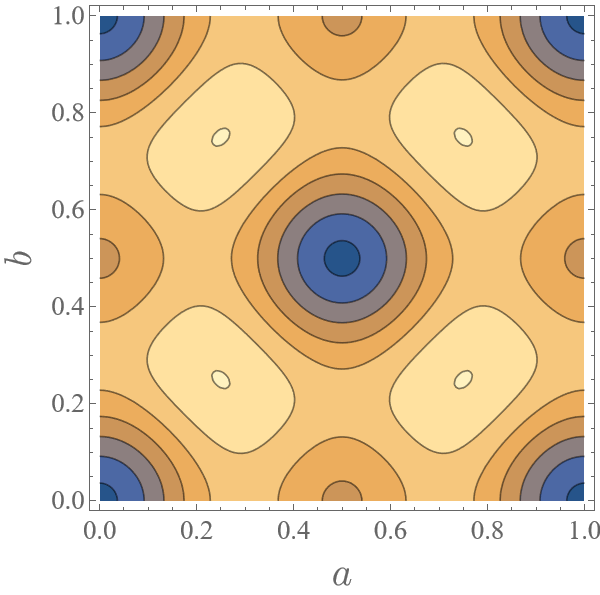} \,
  \includegraphics[width=3.9cm,clip]{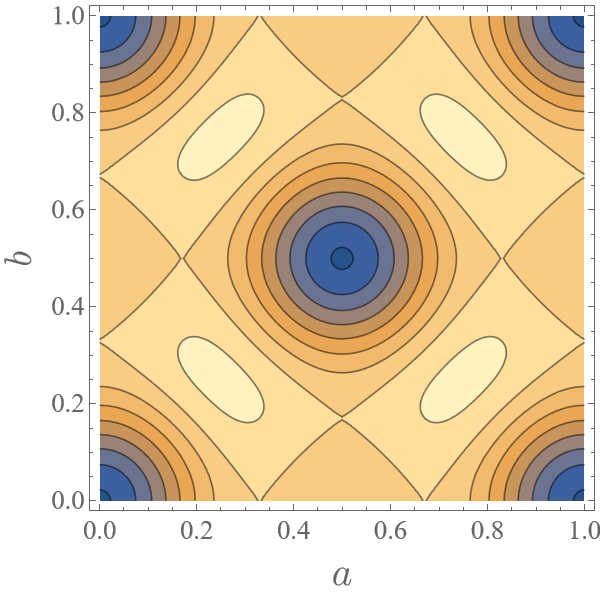} \\
  \includegraphics[width=4cm,clip]{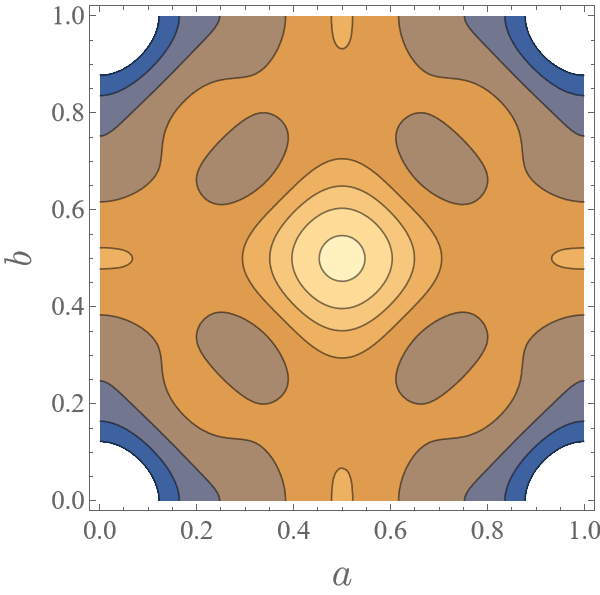} \,
  \includegraphics[width=4cm,clip]{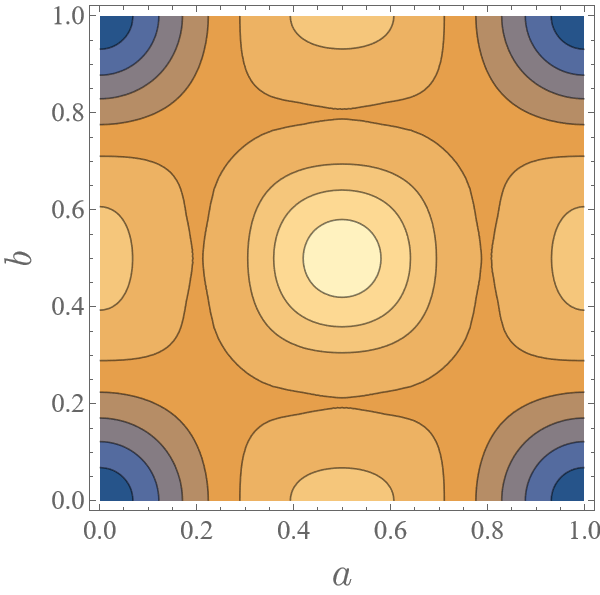} \,
  \includegraphics[width=4cm,clip]{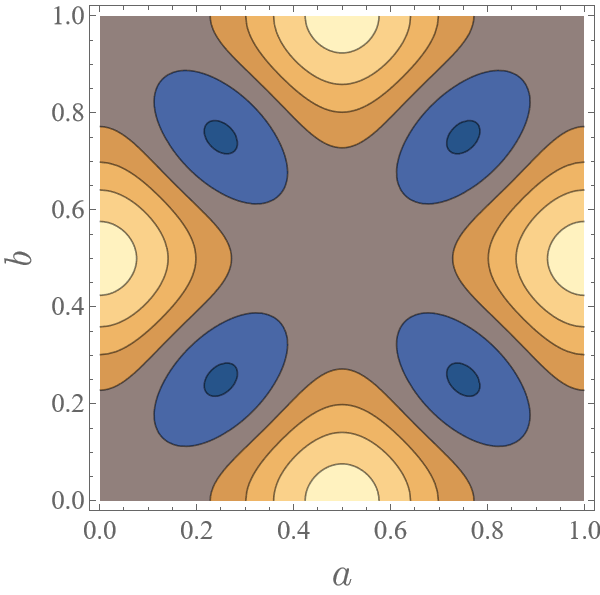} \,
  \caption{
    In the upper row, the figures depict the contributions from a bosonic d.o.f. in \repr6, 
    \repr{15}, \repr{21} and \repr{35}, respectively, from left to right, while the 
    figures in the lower row depict the contributions from a d.o.f. in \repr{56}, \repr{70} 
    and \repr{20} for $\beta_T=0$.
    From the light orange region to the dark blue region, values of the potential decrease.}\bigskip
\label{fig:V_R+}
\end{figure}
\begin{figure}[]
\centering
  \includegraphics[width=3.9cm,clip]{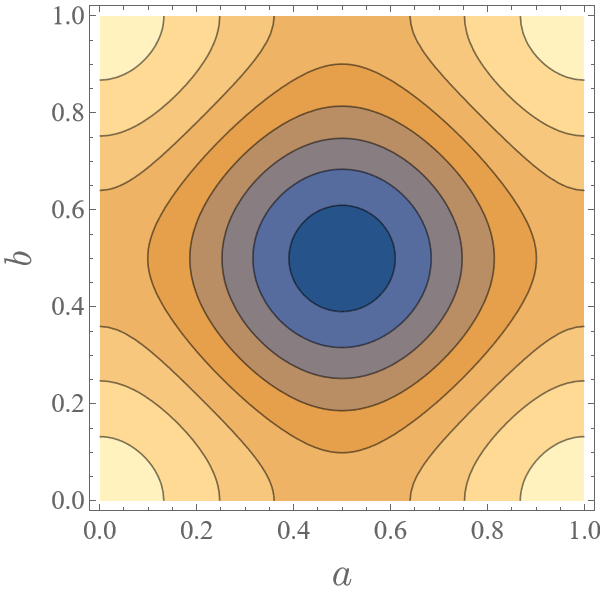} \,
  \includegraphics[width=3.9cm,clip]{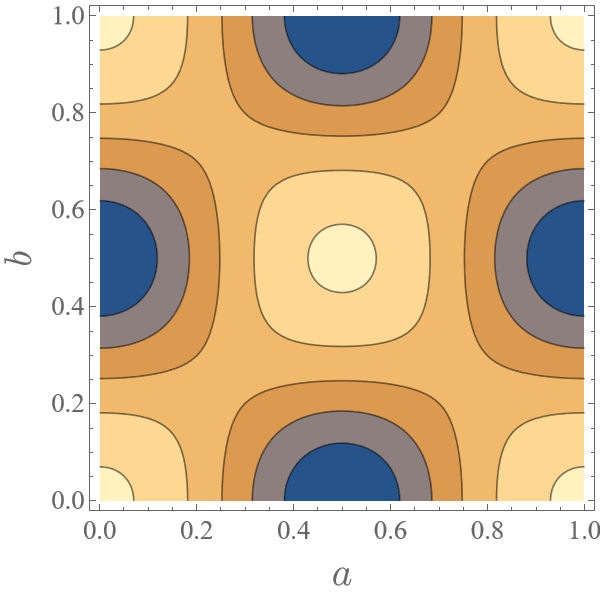} \,
  \includegraphics[width=3.9cm,clip]{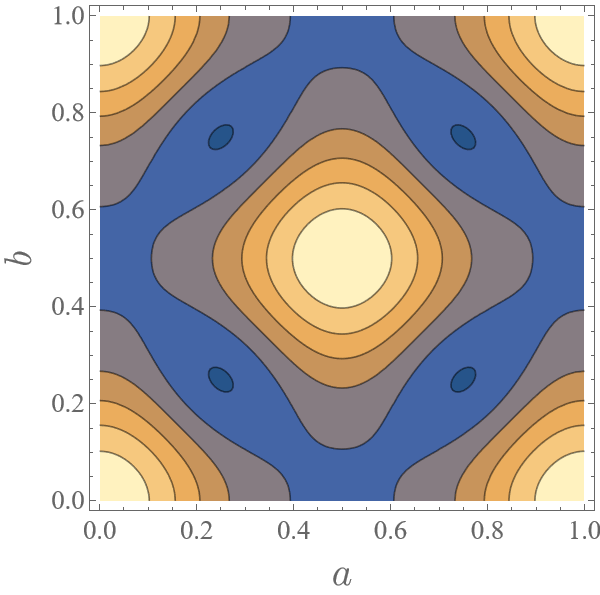} \,
  \includegraphics[width=3.9cm,clip]{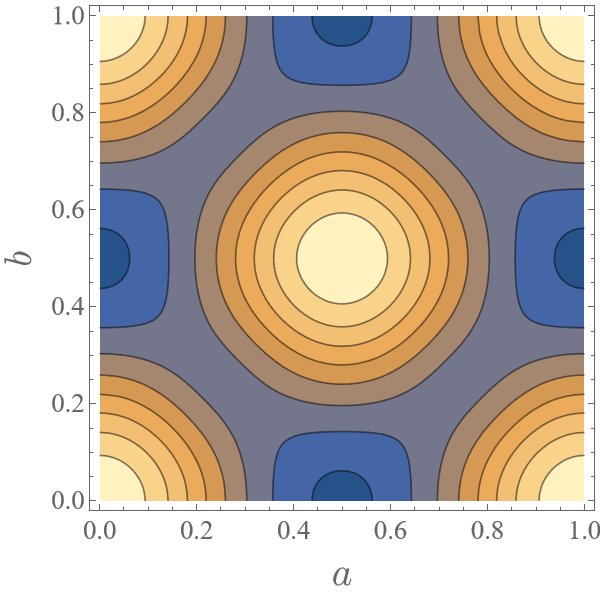} \\
  \includegraphics[width=4cm,clip]{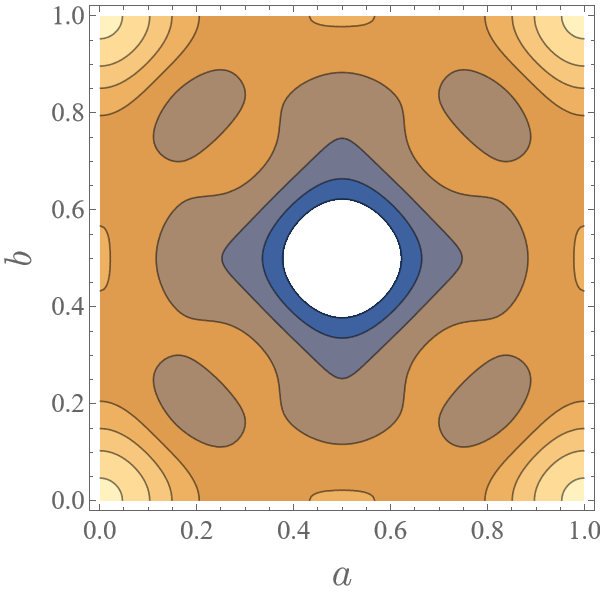} \,
  \includegraphics[width=4cm,clip]{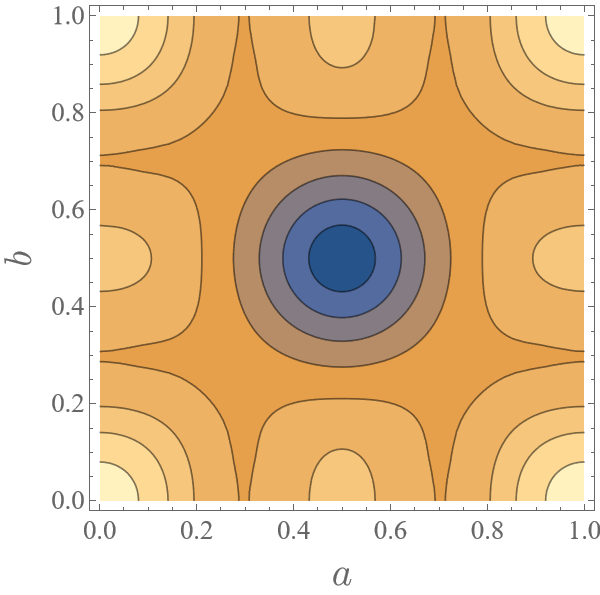} \,
  \includegraphics[width=4cm,clip]{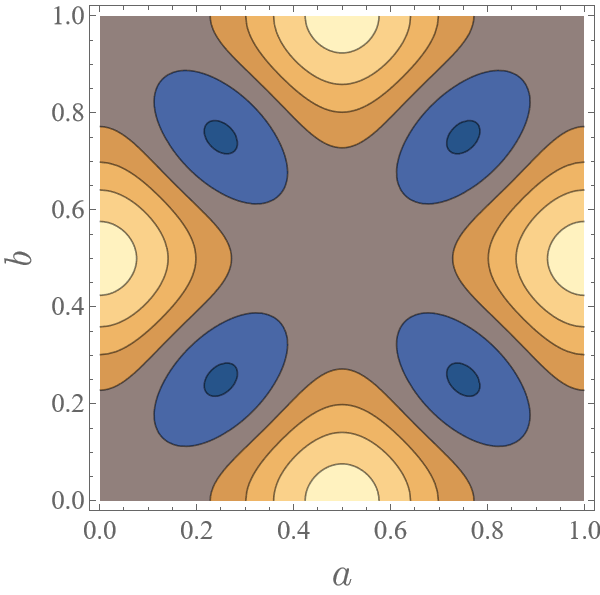} \,
  \caption{
    In the upper row, the figures depict the contributions from a bosonic d.o.f. in \repr6, 
    \repr{15}, \repr{21} and \repr{35}, respectively, from left to right, while the 
    figures in the lower row depict the contributions from a d.o.f. in \repr{56}, \repr{70} 
    and \repr{20} for $\beta_T=1/2$.
    From the light orange region to the dark blue region, values of the potential decrease.}\bigskip
\label{fig:V_R-}
\end{figure}
In figures \ref{fig:V_R+} and \ref{fig:V_R-}, the contributions in eqs.~(\ref{Eq:V-6})--(\ref{Eq:V-20}) are shown for $\beta_T=0,1/2$, respectively.
Among them, those from a bosonic d.o.f. in \repr{15} for $\beta_T=0$ and \repr{20} have the global maximum at the point $(a,b)=(0,1/2)$ (and those equivalent to it). Thus, in models where the contributions from the fermions with $\eta_T=+1$ in \repr{15} or \repr{20} representations dominate the effective potential, the analyses around this vacuum are justified. 

\begin{figure}[]
\centering
  \includegraphics[height=6cm,clip]{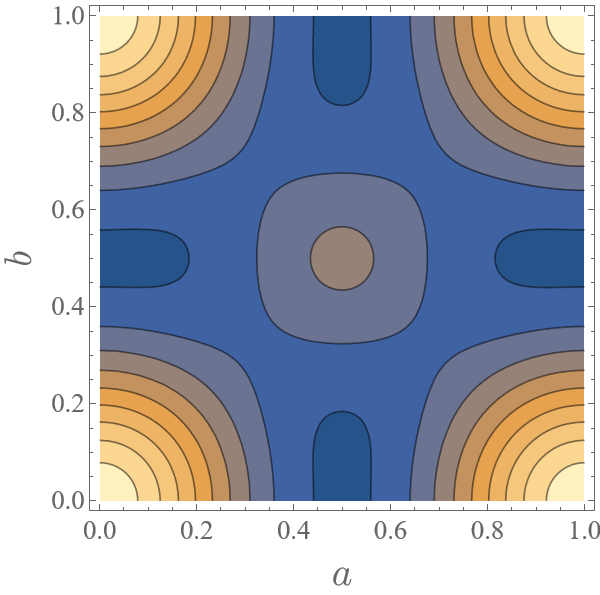} \qquad
  \includegraphics[height=6cm,clip]{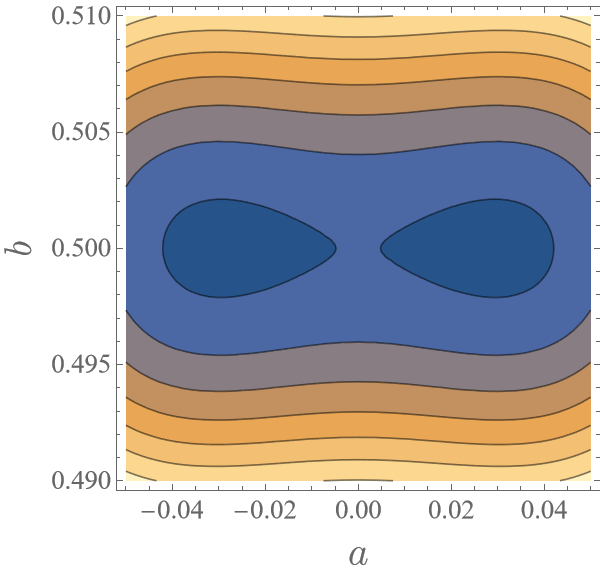} 
  \caption{
    The effective potential in the toy model is depicted. 
    From the light orange region to the dark blue region, values of the potential decrease.}\bigskip
\label{fig:V_EWSB}
\end{figure}

Although we may expect that the further EWSB would be triggered by the large top Yukawa coupling in realistic models, it is possible to construct (toy) models only with massless fields that realize the EWSB. For instance,
if we introduce an adjoint chiral fermion and two Dirac fermions, one in \repr6 and the other in \repr{15}, all with $\eta_T=+1$ as bulk matter fields,
we obtain the effective potential shown in Figure~\ref{fig:V_EWSB}.
In this case, 
the global minimum appears at $(a,b)=(0.0294,1/2)$, and  the compactification scale is estimated as $1/R= m_{\rm W}/0.0294\sim \textrm{a few TeV}$, where $m_{\rm W}$ is the mass of the W boson.
We note that this toy model requires some additional mechanisms to cancel the bulk anomaly.\footnote{
The bulk anomaly explicitly vanishes
    if we introduce an adjoint chiral fermion and chiral fermions having opposite chirality to the adjoint, 16 pieces of \repr6 and 2 pieces of \repr{15}, as bulk matter fields~\cite{Bhardwaj:2015xxa}.
Thus, there is an anomaly-free model that has the same potential as Figure~\ref{fig:V_EWSB} except for an overall factor, whose bulk matter contents consist of an adjoint chiral fermion, 16 Weyl fermions in \repr6, two Weyl fermions in \repr{15}, and seven Dirac fermions in \repr{15}, all with $\eta_T=+1$. 
}
In addition, as suggested above, this model does not reproduce the realistic Yukawa couplings,
particularly the top Yukawa coupling. Thus, the position of the minimum should not be taken seriously.

%
\section{Two Higgs doublet model}
\label{sec:HeavierHiggs}
%

In this section, we aim to understand the above results in terms of the usual two Higgs doublet model. 

In the basis shown in \Eqref{Eq:SU(6)GHU-R0W1}, the zero modes that are identified with the Higgs fields $H_u$ and $H_d$, indicated in eqs.~(\ref{Eq:SU(6)GHU-Hu}) and (\ref{Eq:SU(6)GHU-Hd}) respectively, reside in the $6\times6$ matrix of $A_z$ ($=A_{\bar z}^\dagger$) as, up to normalization,
\bequ
 A_z
   \ni \frac1{\sqrt2}
       \left(\begin{array}{c:c|cc}
         &&H_u& \\ \hdashline
         &&&H_u \\ \hline
         &H_d^T&& \\
         H_d^T&&&
      \end{array} \right), \qquad
  A_{\bar z}
   \ni \frac1{\sqrt2}
       \left(\begin{array}{c:c|cc}
         &&&H_d^* \\ \hdashline
         &&H_d^*& \\ \hline
         H_u^\dagger&&& \\
         &H_u^\dagger&&
      \end{array} \right).
\label{Eq:SU(6)GHU-HuHd}
\eequ

In this basis, the continuous Wilson line phases $W_1$ around the EW symmetric vacuum $(a,b)=(0,1/2)$ are written as
\begin{align}
\notag
W_1
&=\left(\begin{array}{cc:cc|cc}
 0 & 0 & 1 & 0 & 0 & 0 \\
 0 & \frac{1}{2} (c_{2 \pi  \delta a}-c_{2 \pi  \delta b}) & 0 & \frac{1}{2} (c_{2 \pi  \delta a}+c_{2 \pi  \delta b}) & \frac{1}{2} (s_{2 \pi  \delta b}+i s_{2 \pi  \delta a}) & \frac{1}{2} i (s_{2 \pi  \delta a}+i s_{2 \pi  \delta b}) \\ \hdashline
 1 & 0 & 0 & 0 & 0 & 0 \\
 0 & \frac{1}{2} (c_{2 \pi  \delta a}+c_{2 \pi  \delta b}) & 0 & \frac{1}{2} (c_{2 \pi  \delta a}-c_{2 \pi  \delta b}) & \frac{1}{2} i (s_{2 \pi  \delta a}+i s_{2 \pi  \delta b}) & \frac{1}{2} (s_{2 \pi  \delta b}+i s_{2 \pi  \delta a}) \\\hline
 0 & \frac{1}{2} i (s_{2 \pi  \delta a}+i s_{2 \pi  \delta b}) & 0 & \frac{1}{2} (s_{2 \pi  \delta b}+i s_{2 \pi  \delta a}) & \frac{1}{2} (c_{2 \pi  \delta a}-c_{2 \pi  \delta b}) & \frac{1}{2} (c_{2 \pi  \delta a}+c_{2 \pi  \delta b}) \\
 0 & \frac{1}{2} (s_{2 \pi  \delta b}+i s_{2 \pi  \delta a}) & 0 & \frac{1}{2} i (s_{2 \pi  \delta a}+i s_{2 \pi  \delta b}) & \frac{1}{2} (c_{2 \pi  \delta a}+c_{2 \pi  \delta b}) & \frac{1}{2} (c_{2 \pi  \delta a}-c_{2 \pi  \delta b}) 
             \end{array} \right)\\
 &       \\
 &     \simeq \left(\begin{array}{cc:cc|cc}
                 &&1&&&\\
                 &&&1&\pi i(\delta a-i\delta b)&\pi i(\delta a+i\delta b) \\ \hdashline
                 1&&&&&\\
                 &1&&&\pi i(\delta a+i\delta b)&\pi i(\delta a-i\delta b) \\ \hline
                 &\pi i(\delta a+i\delta b)&&\pi i(\delta a-i\delta b)&&1\\
                 &\pi i(\delta a-i\delta b)&&\pi i(\delta a+i\delta b)&1&
              \end{array} \right), 
\label{Eq:SU(6)GHU-W1w/ab}
\end{align}
where $c_\theta= \cos\theta$ and $s_\theta= \sin\theta$. Here, $\delta a$ and $\delta b$ are fluctuations around the vacuum defined as $(a,b)=(\delta a,1/2+\delta b)$, and their higher order terms are neglected in eq.~\eqref{Eq:SU(6)GHU-W1w/ab}.  These fluctuations are observed as 4D fields, in the present case as components of the Higgs doublets $H_u$ and $H_d$.  Apart from the charged components, their neutral components, $H_u^0$ and $H_d^0$, have four real d.o.f. While one of them is gauged away, there is one physical
d.o.f. that does not appear in the continuous Wilson line phases.  This degree corresponds to the nonflat direction, along which the tree-level potential is lifted. To be more concrete, we can write the neutral components as
  \bequ 
  H_u^0=\frac1{\sqrt2}h_ue^{i\phi_u},\qquad
  H_d^0=\frac1{\sqrt2}h_de^{i\phi_d},
\label{Eq:neutralHiggsSV}
\eequ  
where $h_u,\phi_u,h_d$ and $\phi_d$ are real fields. By a gauge transformation, we can set $\phi_u=\phi_d=\phi$. Along the flat directions, $h_u=h_d$ holds. In other words, one of the flat directions corresponds to the linear combination $h_+$ of $h_\pm=(h_u\pm h_d)/\sqrt2$, which leads to $h_u=(h_++h_-)/\sqrt2$ and $h_d=(h_+-h_-)/\sqrt2$, and $h_-$ is set to 0. We note that it is just an approximation and in general $h_-$ appears as a physical d.o.f.

Next, let us examine mass eigenstates. For this purpose, we should work on a broken vacuum. An example is the one at the end of section~\ref{sec:EffPot}, shown in Figure~\ref{fig:V_EWSB}, where $\delta a$ acquires a small VEV. It corresponds to the case where $h_+$ gets a small VEV so that $h_+=v+h$ where $h$ is the fluctuation along the flat direction. Around this vacuum, the neutral components are written as
\bequ
 H_u^0=\frac12(v+h+h_-+iA),\qquad
 H_d^0=\frac12(v+h-h_-+iA),
\label{Eq:neutralHiggsBV}
\eequ
where $A=v\phi$ and higher order terms in the fluctuations are neglected. As far as $v$ is small enough, the effects of the quartic terms are negligible in the estimation of the masses, except for the lightest mode, or the SM-like Higgs. Thus, approximating that the lightest mode is massless, we consider the general mass terms, 
\bequ
 {\cal L}_2=m_{H_u}^2\abs{H_u}^2+m_{H_d}^2\abs{H_d}^2+(BH_uH_d+{\rm h.c.}).
\label{Eq:generalMassIn2HDM}
\eequ 
A straightforward calculation leads to 
\bequ
 {\cal L}_2=\half m_h^2(h+v)^2+\half M_A^2(h_-^2+A^2)+\frac{m_{H_u}^2-m_{H_d}^2}2(h+v)h_-
            -A(h+v){\rm Im} B,
\label{Eq:L2}
\eequ 
where the charged components are suppressed and 
\bequ
 m_h^2=\frac{m_{H_u}^2+m_{H_d}^2}2+{\rm Re}B,\qquad
 M_A^2=\frac{m_{H_u}^2+m_{H_d}^2}2-{\rm Re}B.
\label{Eq:mhMA}
\eequ
The calculation in section~\ref{sec:EffPot} is carried out with $h_-=0$, in toy models where $H_u$ and $H_d$ stand on equal footing and there are no phases, leading to $m_{H_u}^2-m_{H_d}^2={\rm Im}B=0$. As for $m_h^2$ and $M_A^2$, these correspond to the eigenvalues of the Hessian matrix around the vacuum up to a proportional constant, within the approximation $m_h^2$ is very small. 

In realistic models, the situation will not be so simple. As mentioned above, the difference between the top and bottom quark masses requires a difference between couplings of $H_u$ and of $H_d$, and/or $\tan\beta=\VEV{h_u}/\VEV{h_d}\neq1$. Then, most likely $m_{H_u}^2$ is different from $m_{H_d}^2$, and the third term in the right-hand side of eq.~\eqref{Eq:L2} is nonzero to make the VEV of $h_-$ nonvanishing. 
To analyze the VEV, the effective potential along the nonflat direction should be calculated. 
For this purpose, the discussion in ref.~\cite{Kojima:2008ky} would be helpful, where 
the one-loop corrections to the mass and the quartic terms for the SUSY partner of the extra-dimensional component of the gauge field, 
namely a zero mode that is not the flat direction in the multiplet with the gauge field, 
are calculated in a 5D toy model. 
However, 
it would not be easy to carry out the calculation in realistic models. Then, it may be reasonable to treat the mass parameters, which are UV sensitive, as free parameters, in view of the low energy effective theory. In this treatment, the so-called gauge-Higgs condition~\cite{Haba:2005kc} will be useful. We will address it in a future work.

%
\section{Conclusions and discussions}
\label{sec:concl}
%

We have studied 6D $SU(n)$ gauge models with rank-reducing discrete BCs on $T^2/\Z4$.  In the absence of continuous Wilson line phases,
product group unification models can be constructed based on an $SU(7)$ gauge model with the BCs of eq.~\eqref{eq:SU(7)-w/oWL-3} and an $SU(8)$ gauge model with the BCs of eq.~\eqref{eq:SU(8)-w/oWL}. 
When continuous Wilson line phases are present,
we have found that a minimal model, namely, an $SU(6)$ gauge model based on the rank-reducing BCs of eq.~\eqref{Eq:SU(6)-w/WL}, can be used as an EW model.
  Thus, the rank-reducing discrete BCs can open new possibilities of phenomenological models.

In particular, the $SU(6)$ model has
several excellent features. 
These BCs allow for the zero modes for the extra-dimensional components of the gauge bosons, which form two Higgs doublets.
In addition, the quarks in each generation can be unified into one multiplet, without exotic quarks, as the zero modes of a bulk field belonging to the $\bm{15}$ representation of $SU(6)$. 

The remaining symmetry of the $SU(6)$ model
in the low energy effective theory depends on the values of the Wilson line phases parameterized by eq.~\eqref{Eq:<Az>}.  For example, the symmetry becomes $U(1)$ for a generic $\alpha_j$, while it is enhanced to $U(1)^4$ for $\alpha_j = 0$ ($j=1,\dots,4$) or $SU(2)_D \times U(1)_y$ for $\alpha_j = -i/4$.  
The last one can be identified with the EW symmetry, and if the effective potential for the continuous Wilson line phases has the global minimum that is slightly distant from the point, a small breaking of the EW symmetry is realized.

We have derived the effective potential for the continuous Wilson line phases written as eq.~\eqref{Eq:V-quartet} contributed from a real bosonic d.o.f. of a $\Z4$ quartet, without mass terms.  Using this formula, we have obtained the contributions from a real bosonic d.o.f. of some $SU(6)$ representations, as listed in eqs.~\eqref{Eq:V-6}--\eqref{Eq:V-20}.  The total effective potential is given by the summation of the above contributions over the bulk fields, with weight factors $2$, $4$ and $-4$ for complex scalars, the gauge field and 6D chiral fermions, respectively.  From the results of numerical calculations, we have found that
the contributions from 
fermions in $\bm{15}$ with $\eta_T=+1$ or $\bm{20}$ representations have the global minimum at the EW symmetric point $\alpha_j = -i/4$, and that a small EWSB can be realized in this model
by adding suitable bulk fields.  For instance, if we introduce an adjoint chiral fermion and two Dirac fermions, one in $\bm{6}$ and the other in $\bm{15}$, all with $\eta_T=+1$ as bulk matter fields, the global minimum appears at $\alpha_j = 0.0147 - i/4$.

It should be noted that the continuous Wilson line phases do not represent all of the d.o.f. of the scalar zero modes.  The calculation only with the continuous Wilson line phases is valid if the VEVs of the other modes are negligible.  In many cases, the condition naturally holds since the effective potential for the other modes is lifted at the tree level, by the commutation relation in the field strength.  In the above case, however, the deviation from the EW symmetric vacuum is also small and the other modes may not be negligible, and they should be taken care of appropriately.  This situation is obvious in the analysis in terms of the two Higgs doublet model, where $\tan\beta$ is generically different from 1, even when there are flat directions along which the quartic term is vanishing.  
Then, the quantum corrections to the effective potential for the other modes 
should be calculated. Since it would be a demanding task in realistic models, which will contain complicated mass terms, the effective theoretical approach would be more efficient and practically effective~\cite{Haba:2005kc}.

In 6D models, in contrast to 5D models, tadpole terms for the field strength may be 
allowed at the fixed points. These tadpole terms generically reintroduce 
quadratically divergent contributions to the Higgs mass term already at the 
one-loop level. In the present analysis, we simply assume that the tadpole 
contributions are sufficiently small

In $M^4 \times T^2/\Z N$ orbifold models, 
an orbifold family unification on the basis of $SU(n)$ gauge theories has been examined, and 
enormous numbers of models
with three families of the SM matters are derived from 
a massless 6D Dirac fermion~\cite{GK&M} 
and a pair of massless 6D Dirac fermions~\cite{G&K}, 
under the BCs with diagonal representatives.
It would be intriguing to study the orbifold family unification
under BCs including non-diagonal twist matrices such as \Eqref{Z4-t'1}.
Furthermore, it is also attractive to construct a grand unified model.  For instance, an $SU(9)$ gauge model with the BCs of eq.~\eqref{eq:SU(9)-w/WL} would be hopeful.
We will address these topics in future works.

\section*{Acknowledgments}
This work was supported in part by scientific grants from the Ministry of Education, Culture, Sports, Science and Technology under Grant No.~22K03632 (YK).

\appendix

%
\section{Kaluza-Klein decompositions}
\label{app:KK}
%

We show the KK decomposition of 6D fields that have no mass terms on $T^2/\mathbb Z_4$ with the BCs specified by the matrices $R_0$ and $W_1$ in eq.~\eqref{Eq:DiagonalW1}, which belong to the fundamental representation of $SU(6)$.  As shown below, under the operation of $R_0$ in eq.~\eqref{Eq:DiagonalW1}, components of 6D fields are classified as quartets, doublets and singlets, among which only quartets can couple with the continuous Wilson line phases.

Let $\Phi(z)$ be a 6D field, an $SU(6)$ multiplet corresponding to an irreducible representation ${\cal R}$ of $SU(6)$.
In addition to the twist matrices $R_0$ and $W_1$ in the fundamental representation in eq.~\eqref{Eq:DiagonalW1}, we denote corresponding twist matrices
  in a representation ${\cal R}$, other than the fundamental representation, by $R_0^{({\cal R})}$ and $W_1^{({\cal R})}$.
The BCs are given by
\begin{align}
  \Phi(z+1)=\eta_TW_1^{({\cal R})}\Phi(z), \qquad 
  \Phi(iz)=\eta_RR_0^{({\cal R})}\Phi(z), 
  \label{Eq:bcgen1}
\end{align}
where $|\eta_T|=|\eta_R|=1$.
The field $\Phi$ is a multiplet and is composed of component fields. Since $W_1$ is diagonalized, component fields can be taken as eigenstates under the BC of the translation in eq.~\eqref{Eq:bcgen1}.  On the other hand, under the BC of the $\pi/2$ rotation in eq.~\eqref{Eq:bcgen1}, component fields are permuted cyclically by $R_0$.

From the above observations, we can write the BCs in eq.~\eqref{Eq:bcgen1} for component fields as
\begin{align}
 \phi_k(z+1)=e^{2\pi i \beta_T}e^{2\pi iq_k}\phi_k(z), \qquad
  \phi_k(iz)=e^{2\pi i\gamma_R}\phi_{k+1}(z), 
  \label{Eq:bcgenforcomp1}
\end{align}
where a component field is denoted by $\phi_k(z)$, and we use $\beta_T$ and $\gamma_R$ to represent $\eta_T$ and $\eta_R$,
respectively.
The d.o.f. of the continuous Wilson line phases are contained in $q_k$. Since the $2\pi$ rotation must be the identity, we require $\phi_{k+4}(z)=\phi_k(z)$, which leads to $4\gamma_R\in \mathbb Z$.\footnote{%
  In a more general setup, phase factors of 
  the $\pi/2$ rotation can depend on $k$ $(k=1,\dots,4)$ as
  $\phi_k(iz)=e^{2\pi i\gamma_k}\phi_{k+1}(z)$, instead of the right
  equation in eq.~\eqref{Eq:bcgenforcomp1}. In this case, the phases
  must satisfy $\gamma_1+\gamma_2+\gamma_3+\gamma_4\in \mathbb
  Z$.}
We can classify component fields obeying the BCs in eq.~\eqref{Eq:bcgenforcomp1} into three types: quartets, doublets and singlets. We define them as follows.  A quartet is composed of four different fields as $\phi_k$ $(k=1,\dots,4)$, where we define $\phi_{k+4}=\phi_k$. A doublet is composed of two different fields as $\phi_k$ $(k=1,2)$, where we define $\phi_{k+2}=\pm \phi_k$.  A singlet is composed of only $\phi_1$, where we define $\phi_{k+1}=i^m\phi_k$ for any $k$ and $m$ $(m\in\mathbb Z)$.

For example, if $\Phi$ belongs to the fundamental representation of $SU(6)$, its BCs in eq.~\eqref{Eq:bcgen1} are written by the matrices in eq.~\eqref{Eq:DiagonalW1}.  From the form of $R_0$ in eq.~\eqref{Eq:DiagonalW1}, one can easily see that $\Phi$ contains a doublet and a quartet.
From the form of $W_1$ in eq.~\eqref{Eq:DiagonalW1}, one also see that
the components of
these doublet and 
quartet have the charges
$(q_1,q_2)=(1/2,0)$ and 
$(q_1,q_2,q_3,q_4)=(b,-a,-b,a)$, respectively.

\subsection{Kaluza-Klein decompositions of quartets}

Let us discuss KK decompositions of quartets $\phi_k$, where $k+4$ is identified with $k$. First, we clarify the constraints for the charges $q_k$ of quartets.  From the BC in the left equation in eq.~\eqref{Eq:bcgenforcomp1}, we obtain
\begin{align}
   \phi_{k+1}(z+1)=e^{2\pi i \beta_T}e^{2\pi iq_{k+1}}\phi_{k+1}(z).
\end{align}
In the above, using the right equation in eq.~\eqref{Eq:bcgenforcomp1}, we can rewrite $\phi_{k+1}$ by $\phi_{k}$, and we obtain the relation $\phi_{k}(z+i)=e^{2\pi i \beta_T}e^{2\pi iq_{k+1}}\phi_{k}(z)$.  From similar discussions, we find a more general formula as 
\begin{align}
  \phi_{k}(z+i^m)=e^{2\pi i \beta_T}e^{2\pi iq_{k+m}}\phi_{k}(z),
  \label{imtrans1}
\end{align}
for $m\in \mathbb Z$. 

The BC in eq.~\eqref{imtrans1} constrains the charges of the quartets. Using the identity $z+i^m+i^{m+2}=z$ with $m=0$, it can be seen that
\begin{align}
  q_k+q_{k+2}+2\beta_T \in \mathbb Z.
  \label{qbc1revchar0}
\end{align}
As discussed in the above, the quartet contained in the fundamental representation of $SU(6)$ has the charges $(q_1,q_2,q_3,q_4)=(b,-a,-b,a)$, which satisfy 
  \begin{align}
  q_k+q_{k+2} \in \mathbb Z. 
  \label{qbc1revchar}
  \end{align}

To discuss the other representations, we note that,
corresponding to eq.~\eqref{qbc1revchar},
the relation $2q_k\in \mathbb Z$ $(k=1,2)$ holds 
for the charges of the doublet in the fundamental representation.
Since the other representations are obtained by the tensor products of the fundamental representation, the charges of a component of an arbitrary representation are given by certain summations of the above charges. 
This implies that, for any representation, the same relation in eq.~\eqref{qbc1revchar}, and thus $2\beta_T \in \mathbb Z$, hold. 
We note that, for any quartet $\phi_k$ $(k=1,\dots,4)$, their charges $q_k$ are not taken independently.

To discuss their KK decomposition, let us introduce mode functions as 
\begin{gather}\label{modedef1}
  f_{n_1+\tilde q_{k},n_2+\tilde q_{k+1}}(z)=
  e^{2\pi i (n_1+\tilde q_{k})y^1}
  e^{2\pi i (n_2+\tilde q_{k+1})y^2},\\
  \label{mc2}
  \bar f_{n_1+\tilde q_{k},n_2+\tilde q_{k+1}}(z)=
  f_{n_1+\tilde q_{k},n_2+\tilde q_{k+1}}(-z)
  =  f_{-n_1-\tilde q_{k},-n_2-\tilde q_{k+1}}(z), 
\end{gather}
where $z=y^1+iy^2$. Here, we have defined
  \begin{align}
    \tilde q_k=q_k+\beta_T,
  \end{align}
for simplicity of the notation. For a fixed $k$, these mode functions are orthogonal and normalized as 
\begin{align}
\iint d^2y\bar f_{n_1'+\tilde q_{k},n_2'+\tilde q_{k+1}}(z)  f_{n_1+\tilde q_{k},n_2+\tilde q_{k+1}}(z)=
\iint d^2y  e^{2\pi i (n_1-n_1')y^1}
  e^{2\pi i (n_2-n_2')y^2}=\delta_{n_1n_1'}\delta_{n_2n_2'}.
\end{align}
Under the discrete rotations, these mode functions are related as 
\begin{align}\label{mc1}
  f_{n_1+\tilde q_k,n_2+\tilde q_{k+1}}(iz)&= f_{n_2+\tilde q_{k+1},-n_1+\tilde q_{k+2}}(z), \\
  f_{n_1+\tilde q_k,n_2+\tilde q_{k+1}}(-z)&= f_{-n_1+\tilde q_{k+2},-n_2+\tilde q_{k+3}}(z), \\
  f_{n_1+\tilde q_k,n_2+\tilde q_{k+1}}(-iz)&= f_{-n_2+\tilde q_{k+3},n_1+\tilde q_{k}}(z),
                                \label{mc3}
\end{align}
where we have used eq.~\eqref{qbc1revchar0}. 

We introduce the KK decomposition of $\phi_k$ that obeys the BCs in eq.~\eqref{Eq:bcgenforcomp1} as
\begin{align}\label{kkplain1}
  \phi_k(z)=\sum_{n_1,n_2\in \mathbb Z}\phi_k^{(n_1+\tilde q_k,n_2+\tilde q_{k+1})}f_{n_1+\tilde q_k,n_2+\tilde q_{k+1}}(z), 
\end{align}
where $\phi_k^{(n_1+\tilde q_k,n_2+\tilde q_{k+1})}$ are 4D fields. To satisfy the right equation of eq.~\eqref{Eq:bcgenforcomp1}, $\phi_k^{(n_1+\tilde q_k,n_2+\tilde q_{k+1})}$ are constrained as
\begin{align}\label{congen3}
  \phi_k^{(-n_2+\tilde q_k,n_1+\tilde q_{k+1})}=e^{2\pi i \gamma_R}\phi_{k+1}^{(n_1+\tilde q_{k+1},n_2+\tilde q_{k+2})}. 
\end{align}
Thus, $\phi_k^{(n_1+\tilde q_k,n_2+\tilde q_{k+1})}$ for $k=2,3,4$ is expressed by $\phi_1^{(n_1+\tilde q_1,n_2+\tilde q_2)}$ as 
\begin{align}\label{congen41}
  \phi_2^{(n_1+\tilde q_2,n_2+\tilde q_3)}&=e^{-2\pi i\gamma_R}\phi_1^{(-n_2+\tilde q_1,n_1+\tilde q_2)}, \\
  \phi_3^{(n_1+\tilde q_3,n_2+\tilde q_4)}&=e^{-2\pi i(2\gamma_R)}\phi_1^{(-n_1+\tilde q_1,-n_2+\tilde q_2)}, \\\label{congen43}
  \phi_4^{(n_1+\tilde q_4,n_2+\tilde q_1)}&=e^{-2\pi i(3\gamma_R)}\phi_1^{(n_2+\tilde q_1,-n_1+\tilde q_2)}.
\end{align}
We can treat $\phi_1^{(n_1+\tilde q_1,n_2+\tilde q_2)}$ as the independent KK modes among 
$\phi_k^{(n_1+\tilde q_k,n_2+\tilde q_{k+1})}$. Let us rewrite the KK decomposition in eq.~\eqref{kkplain1} as follows: 
\begin{align}\label{kkeq411}
  \phi_1(z)&=\sum_{n_1,n_2\in \mathbb Z}\phi_1^{(n_1+\tilde q_1,n_2+\tilde q_2)}f_{n_1+\tilde q_1,n_2+\tilde q_2}(z), \\
  \phi_2(z)&=e^{-2\pi i \gamma_R}\sum_{n_1,n_2\in \mathbb Z}
             \phi_1^{(n_1+\tilde q_1,n_2+\tilde q_2)}  f_{n_2+\tilde q_2,-n_1+\tilde q_3}(z) , \\
  \phi_3(z)&=e^{-2\pi i (2\gamma_R)}\sum_{n_1,n_2\in \mathbb Z}
             \phi_1^{(n_1+\tilde q_1,n_2+\tilde q_2)} f_{-n_1+\tilde q_3,-n_2+\tilde q_4}(z), \\
  \phi_4(z)&=e^{-2\pi i (3\gamma_R)}\sum_{n_1,n_2\in \mathbb Z}
             \phi_1^{(n_1+\tilde q_1,n_2+\tilde q_2)}f_{-n_2+\tilde q_4,n_1+\tilde q_1}(z).
             \label{kkeq414}
\end{align}

Finally, we derive KK masses. As a concrete example, suppose that $\phi_k$ is a complex scalar field. Then, the 6D kinetic term is given by 
\begin{align}\label{massL1}
{\cal L}_1= \iint d^2y \sum_{k=1}^4\phi_k^* (\der_\mu^2-\der_{y^1}^2-\der_{y^2}^2)\phi_k. 
\end{align}
From the KK decompositions in eqs.~\eqref{kkeq411}--\eqref{kkeq414} and
eq.~\eqref{qbc1revchar0}, one finds 
\begin{align}\label{massL1new}
  &  {\cal L}_1
    =4\sum_{n_1,n_2\in\mathbb Z}\phi_1^{(n_1+\tilde q_1,n_2+\tilde q_2)*}(\der_\mu^2+(M_{\rm KK}^2)^{(n_1,n_2)})\phi_1^{(n_1+\tilde q_1,n_2+\tilde q_2)},
\end{align}
where $(M_{\rm KK}^2)^{(n_1,n_2)}$ are the KK masses for $\phi_1^{(n_1+\tilde q_1,n_2+\tilde q_2)}$ and given by 
\begin{align}
\label{kkmasst41}
  (M_{\rm KK}^2)^{(n_1,n_2)}
  =(2\pi)^2\left[
    (n_1+q_1+\beta_T)^2    +(n_2+q_2+\beta_T)^2\right].
\end{align}
Thus, the KK masses for a quartet are determined by $\beta_T$ and the charges $q_1$ and $q_2$.

\subsection{Doublets and singlets have no coupling with continuous Wilson line phases}

Here, we show that the doublets and the singlets have no coupling with the continuous Wilson line phases.  For a doublet (singlet),   $\phi_{k+2}$ $(\phi_{k+1})$ is proportional to $\phi_k$.  The BCs in eq.~\eqref{Eq:bcgenforcomp1} 
lead to the same constraint as in eq.~\eqref{qbc1revchar0} also for the doublets and the singlets, for which $q_{k+2}=q_k$ holds. Thus, eq.~\eqref{qbc1revchar0} and $2\beta_T\in \mathbb Z$ mean $2q_k\in \mathbb Z$. These fields cannot couple with the continuous Wilson line phases since the charges $q_k$ of the doublets and the singlets are discretized.  While we do not show explicitly, their KK expansions and KK masses can be derived as in the case of the quartets.

%
\section{Derivation of effective potentials}
\label{app:epot}
%

In this appendix, we derive the one-loop effective potential for the continuous Wilson line phases.  As shown in appendix~\ref{app:KK}, only quartets couple with the continuous Wilson line phases. For bulk complex scalars that have no mass terms, their 4D effective Lagrangian density and KK masses are given by eqs.~\eqref{massL1new} and~\eqref{kkmasst41}. A similar discussion for the case with a bulk 6D Weyl fermion that has no mass terms gives the same KK mass. Functional integrations yield the
following contribution to the effective potential:
\begin{align}
  \hat N {\cal V}^{[\beta_T]}(q_1,q_2)={\hat N\over 2}\sum_{n_1,n_2\in \mathbb Z}
  \int {d^4p_{\rm E}\over (2\pi)^4}\ln\left[p_{\rm E}^2+(M_{\rm KK}^2)^{(n_1,n_2)}\right],
  \label{vcont1}
\end{align}
where we have introduced $\hat N$ for later convenience. We define $\hat N=2$ $(-4)$ for the case with a bulk complex scalar (6D Weyl fermion).

The contribution to the effective potential in eq.~\eqref{vcont1} is evaluated by a standard procedure that employs the zeta function regularization. 
After the momentum integral, we have 
\begin{align}
   {\cal V}^{[\beta_T]}(q_1,q_2)=-{1\over 32\pi^2}\sum_{n_1,n_2\in \mathbb Z}
  \int_0^\infty dt t^{-3}e^{-(M_{\rm KK}^2)^{(n_1,n_2)}t}. 
  \label{vcont2}
\end{align}
Here, the Poisson resummation formula, 
\begin{align}\label{poissongen}
  \sum_{\bm n}e^{-\pi(n_i+d_i)(A^{-1})_{ij}(n_j+d_j)}=\sqrt{\det A}\sum_{\bm w} e^{-\pi w_iA_{ij}w_j}e^{2\pi i w_kd_k}, 
\end{align}
can be used. In general, $\bm n$ and $\bm w$ are $D$-dimensional vectors, whose components are expressed by $n_i$ and $w_i$. The summations are taken over all integers. In the present case, the 2D formula with 
\begin{gather}
 A=(4\pi t)^{-1}{\rm diag}(1,1), \qquad 
  d_1=q_1+\beta_T, \qquad d_2=q_2+\beta_T, 
\end{gather}
is suitable for eq.~\eqref{vcont2}. Then, we obtain 
\begin{align}\label{efpotgen1}
   {\cal V}^{[\beta_T]}(q_1,q_2)&=
            -{1\over 128\pi^3}\sum_{w_1,w_2\in \mathbb Z}
            e^{2\pi i (w_1(q_1+\beta_T)+w_2 (q_2+\beta_T))}
            \int_0^\infty dt t^{-4}
            e^{-{1\over 4t}(w_1^2+w_2^2)}\\
  &=-{1\over\pi^3}\sum_{(w_1,w_2)\neq (0,0)}
    {e^{2\pi i (w_1(q_1+\beta_T)+w_2 (q_2+\beta_T))}\over
    ( w_1^2+w_2^2)^3}+({\rm const.}).
    \label{effpotform1}
\end{align}
In the last line, we have separated the term corresponding to $(w_1,w_2)=(0,0)$, which is independent of the continuous Wilson line phases and thus is an irrelevant constant in the effective potential.

From the expression in eq.~\eqref{effpotform1}, it can be seen that the effective potential has various symmetries with respect to changes of $(q_1,q_2)$. For example, integer shifts, sign flips and the exchange between $q_1$ and $q_2$ keep the effective potential contribution. 

In this model, we can find quartets for a given $SU(6)$ multiplet, and their charges $q_1$ and $q_2$ can be specified. Then, using the formula in eq.~\eqref{effpotform1}, we can evaluate effective potential contributions as discussed in section~\ref{sec:EffPot-repr}.


\end{document}